\documentclass[aps,prx,twocolumn,superscriptaddress]{revtex4}

\usepackage{amsmath,amssymb,graphicx,xspace,color}

\begin{document}

\title{Spectroscopic evidence of nematic fluctuations in LiFeAs}

\author{Zhixiang Sun}

\affiliation{IFW Dresden, 01069 Dresden, Germany}

\author{Pranab Kumar Nag}

\affiliation{IFW Dresden, 01069 Dresden, Germany}

\author{Steffen Sykora}

\affiliation{IFW Dresden, 01069 Dresden, Germany}

\author{Jose M. Guevara}

\affiliation{IFW Dresden, 01069 Dresden, Germany}

\author{Sven Hoffmann}

\affiliation{IFW Dresden, 01069 Dresden, Germany}

\author{Christian Salazar}

\affiliation{IFW Dresden, 01069 Dresden, Germany}

\author{Torben H\"{a}nke}

\affiliation{IFW Dresden, 01069 Dresden, Germany}

\author{Rhea Kappenberger}

\affiliation{IFW Dresden, 01069 Dresden, Germany}

\affiliation{Institute for Solid State Physics, TU Dresden, 01069 Dresden, Germany}

\author{Sabine Wurmehl}

\affiliation{IFW Dresden, 01069 Dresden, Germany}

\affiliation{Institute for Solid State Physics, TU Dresden, 01069 Dresden, Germany}

\author{Bernd B\"{u}chner}

\affiliation{IFW Dresden, 01069 Dresden, Germany}

\affiliation{Institute for Solid State Physics, TU Dresden, 01069 Dresden, Germany}

\affiliation{Center for Transport and Devices, TU Dresden, 01069 Dresden, Germany}

\author{Christian Hess}

\email[]{c.hess@ifw-dresden.de}

\affiliation{IFW Dresden, 01069 Dresden, Germany}

\affiliation{Center for Transport and Devices, TU Dresden, 01069 Dresden, Germany}

\date{\today}

\begin{abstract}

The role of nematic fluctuations in  the  pairing mechanism  of iron-based superconductors is frequently debated. Here we present a novel method to reveal such fluctuations by identifying
energy and momentum of
the corresponding nematic boson 
through the detection of a boson-assisted resonant amplification of Friedel oscillations. Using Fourier-transform scanning tunneling spectroscopy, we observe
for the unconventional superconductor LiFeAs strong signatures of bosonic states at momentum $q\sim 0$ and energy $\Omega\approx8$~meV. We show that these bosonic states survive in the normal conducting state, and, moreover, that they are in perfect agreement with well-known strong above-gap anomalies in the tunneling spectra. Attributing  these small-$q$  boson modes to nematic fluctuations we provide  the first spectroscopic approach to the nematic boson in an unconventional superconductor.

\end{abstract}


% \pacs{}


%\keywords{}

\maketitle

%%%%%%%%%%%%%%%%%%%%%INTRODUCTION%%%%%%%%%%%%%%%%%%%%%%%%

\section{Introduction}
\label{Sec_Introduction}

The identification of the fine structure of tunneling spectra of strong-coupling conventional superconductors with the fingerprints of the phononic Cooper pairing glue counts as a fundamental step in the rationalization of superconductivity \cite{McMillan1965,Scalapino1966}.
The extension of this approach to unconventional superconductors, such as cuprates and iron-based superconductors (IBS) is highly desirable for clarifying the nature of superconductivity in these materials. 
However, despite of salient above-gap anomalies often present in tunneling spectra \cite{Hudson1999,Jenkins2009,Wang2013d,Song2014,Chi2012,Nag2016}, their interpretation typically remains elusive. A major reason, apart from difficulties to differentiate between elastic and inelastic tunneling contributions \cite{Hlobil2017}, is the lack of momentum information. An accurate resolution of spectral properties in momentum space
is, however, crucial for rationalizing superconductivity in multi-band materials such as IBS.

In many canonical IBS superconductivity emerges upon doping from an antiferromagnetic spin-density-wave (SDW) parent state which probably is related with Fermi surface nesting \cite{Johnston2010}. The SDW state furthermore seems intimately connected with unidirectional electronic, so-called nematic order, involving orbital degrees of freedom \cite{Fernandes2014}.
These proximities of superconductivity and antiferromagnetic order on the one hand and nematic order on the other,
have nourished pertinent scenarios for the mechanism of the Cooper pairing, i.e., respectively, antiferromagnetic spin fluctuations \cite{Mazin2008} and orbital fluctuations \cite{Kontani2010} are conjectured to drive the superconductivity in the IBS.

A strong antiferromagnetic spin resonance, which would be supportive of the spin fluctuation scenario, has been detected in inelastic neutron scattering for some of the prototypical IBS \cite{Inosov2010a}.
While similar spectroscopic signatures of nematic fluctuations supporting superconductivity, however, do not exist up to present, evidence for the relevance of small-$q$ nematic fluctuations for superconductivity has been accumulating \cite{Fernandes2014}. Prominent recent examples are \textit{static} small-$q$ electronic density variations observed in tunneling experiments on FeSe thin films \cite{Li2017} and strained LiFeAs \cite{Yim2018}, where superconductivity is suppressed. Theoretically, the influence of \textit{dynamic} nematic fluctuations can be modeled by a coupling of itinerant electrons to Ising nematic bosons within the framework of an Eliashberg treatment \cite{Lederer2015}. These small-momentum nematic modes play a role similar to that of phonons in a conventional superconductor with the difference that the pairing potential becomes strongly momentum-dependent, is attractive in all pairing channels, and so enhances $T_c$ \cite{Lederer2015}. A generalization of this model to a system of coupled fermion bond density and pseudospin-1/2 degree of freedom has been solved by Quantum Monte Carlo simulations \cite{Schattner2016} and the enhancement of superconducting pairing has been confirmed.

\begin{figure}[t]

\centering

\includegraphics[width=\columnwidth]{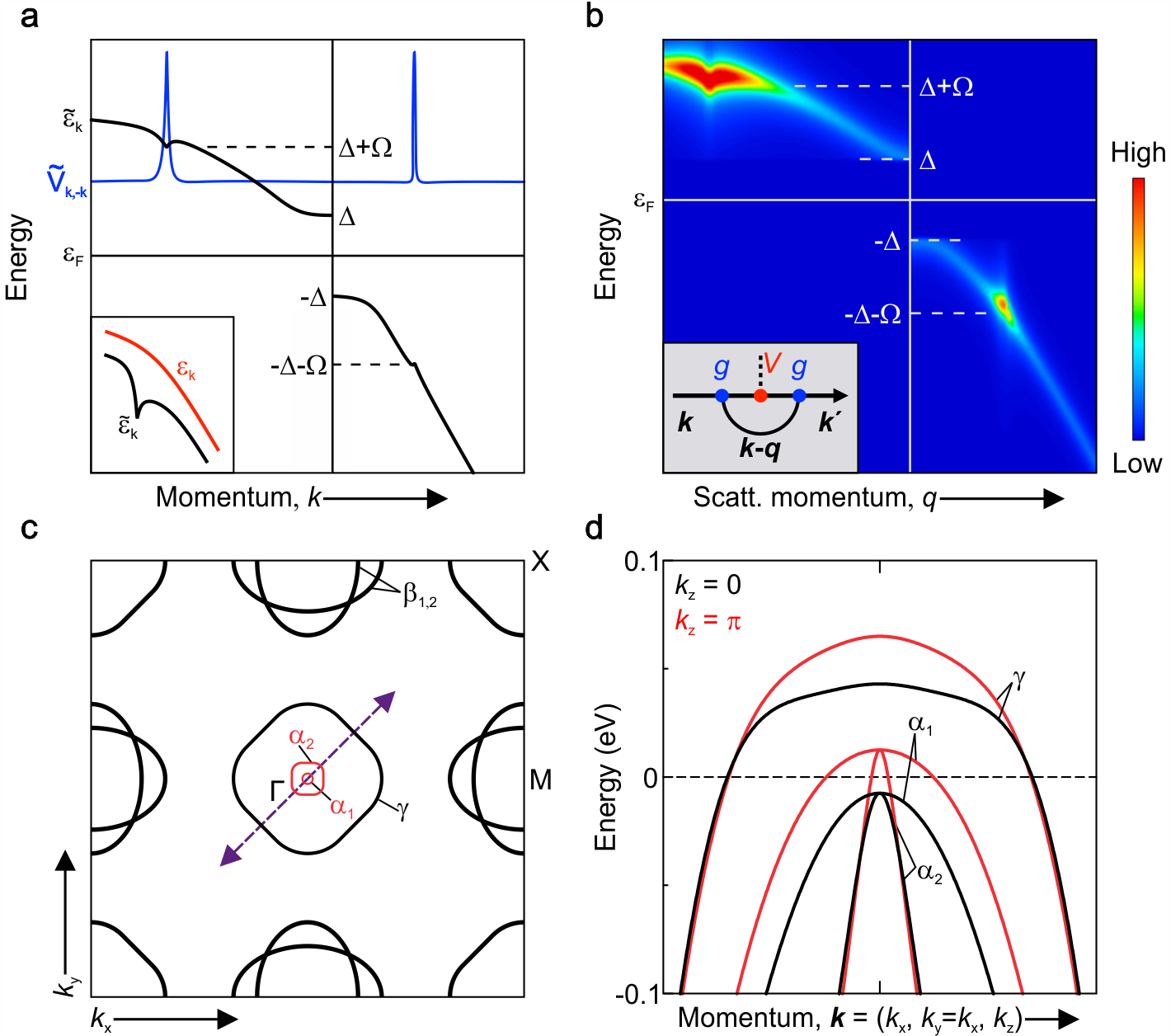}

\caption{Illustration of the boson-enhanced QPI and of the fermiology of LiFeAs. a) Sketch of a hole-like band $\tilde{\varepsilon}_{\bf k}$ (black) in the superconducting state which is renormalized due to electron-boson coupling giving rise to the well-known kink-like structures above and below the Fermi level at energies indicated by dashed lines (the inset shows the bare band in red). Blue line: renormalized impurity potential $\tilde{V}_{{\bf k},-{\bf k}}$ combining opposite momentum vectors representing relevant elastic scattering processes in the QPI. The renormalization is strong at the particular ${\bf k}$ points of the kink structure. b) Sketch of the Fourier-transformed LDOS arising from the renormalized quantities $\tilde{\varepsilon}_{\bf k}$ and $\tilde{V}_{{\bf k},-{\bf k}}$ of (a) (see Appendix~\ref{Subsec_theory}). The intensity is strongly enhanced around the particular scattering momentum combining points where both the fermion band as well as the impurity scattering potential are strongly renormalized due to electron-boson coupling. Inset: Feynman diagram of the corresponding dominant scattering process which is dressed by the excitation of a virtual bosonic mode. c) Illustration of the FS of LiFeAs \cite{Wang2013}. Black: projection of quasi two-dimensional FS pockets to the $k_z=0$ plane. Red: FS pockets of the $\alpha$-bands at $k_z=\pi$ ($Z$-point) which appear only near this point. d) Dispersion of the hole-like bands along the dashed arrow in (c) for $k_{z}=0$ (black) and $k_{z}=\pi$ (red).
}

\label{FS_schem}

\end{figure}

Here we report for the first time spectroscopic evidence of such small-momentum bosonic modes representing nematic fluctuations. To this end, we exploit
a new combined theoretical and experimental approach for detecting the signatures of 
bosonic degrees of freedom in quantum materials %that may contribute to the pairing 
using Fourier transform scanning tunneling spectroscopy (FT-STS) experiments. FT-STS is well established to detect the momentum space representation of the so-called quasiparticle interference (QPI), i.e. the Fourier transform of real-space wave-like modifications of the local density of states (LDOS) caused by an impurity, i.e. the Friedel oscillations. 
The \textit{geometry} of the Friedel oscillations in momentum space has successfully been used to reconstruct the electronic band structure of many correlated materials \cite{Hoffman2002,Aynajian2012,Allan2012,Haenke2012,Hess2013,Grothe2013,Allan2015,Wang2017}. This includes, since very recently, even the detection of subtle band renormalizations due to electron-boson interactions \cite{Grothe2013,Allan2015,Wang2017}. Despite this enormous success of QPI analysis,
the impact of the electron-boson interaction on the Friedel oscillation itself has remained largely unexplored. We have investigated this aspect theoretically (see Appendix~\ref{Subsec_theory} for details) and find that
the impurity scattering potential and thus the \textit{amplitude} of the QPI is resonantly enhanced if the involved electronic states are interacting with a boson, see Figs.~\ref{FS_schem}(a), (b). The effect is strong and implies that an amplitude analysis of the QPI can yield signatures of an interacting boson, including specific information about the bosonic momentum and energy.
The exploitation of this effect in tunneling experiments can therefore be viewed as the addition of momentum information to the analysis of bosonic signatures in tunneling spectroscopy.
We particularly point out that our method can be used to detect a boson independently of its nature, i.e. the boson could be the effective representation of spin, charge, orbital, or nematic fluctuations, or it describes phonons.
The mentioned amplitude sensitivity is here exploited to investigate the unconventional superconductor LiFeAs, where the method works particularly well, as we will show below.

LiFeAs differs in its properties from most other IBS for the following reasons: It is a stoichiometric superconductor which shows no sign of Fermi surface nesting \cite{Borisenko2010,Zeng2013} and no magnetic or nematic order, even under doping \cite{Aswartham2011a,Pitcher2010}. 
Instead of an antiferromagnetic spin resonance, only weak signatures of spin excitations are observed at incommensurate positions in momentum space \cite{Qureshi2012}
which are understood to arise from inter-band transitions between the quasi two-dimensional large hole-like and the electron-like Fermi surface (FS) pockets (labeled $\gamma$ and $\beta$, respectively, see Fig.~\ref{FS_schem}(c)) \cite{Knolle2012}. In fact, these weak spin fluctuations exhibit only subtle changes upon switching between the normal and superconducting state which renders these fluctuations poor candidates for providing the superconducting pairing interaction.

\begin{figure*}[t]

\centering           

\includegraphics[width=0.9\textwidth]{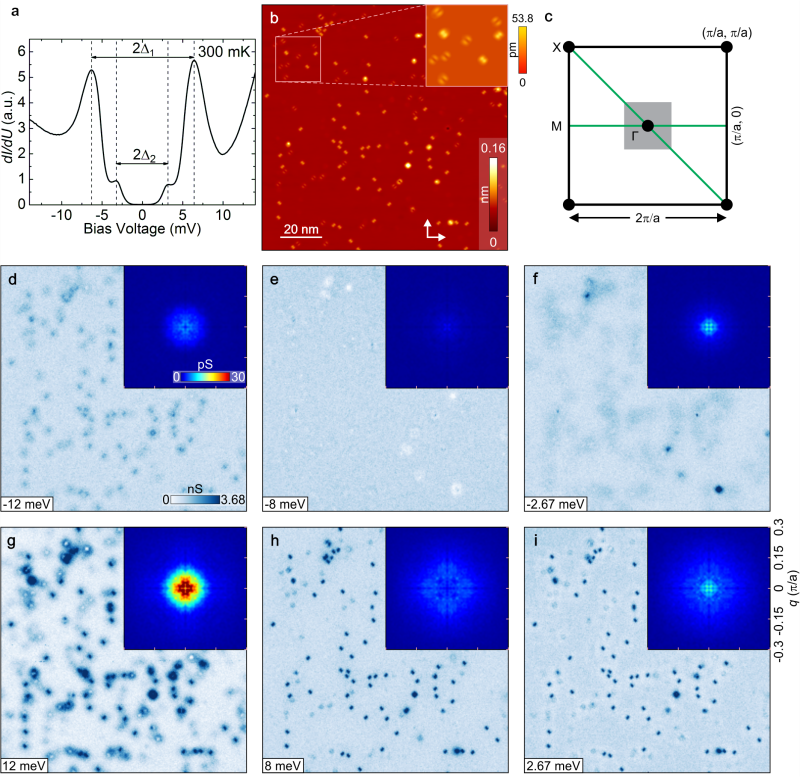}

\caption{Experimental spectroscopic tunneling data in the superconducting state. a) Average $\mathrm{d}I/\mathrm{d}U$ spectrum measured on a defect free surface area of LiFeAs at $T=300$~mK. b) A representative surface topography ($U_\mathrm{bias} = -50$~mV, $I = 100$~pA) with clearly identifiable Fe-defects (inset). The directions of the shortest Fe-Fe distance $a=2.68$~{\AA} \cite{Pitcher2010} are indicated by arrows. c) Illustration of the space of in-plane scattering vectors. Green lines indicate the high-symmetry directions considered in Fig.~\ref{Fig_3}. The gray square shows the $q$-space area covered by the insets in panels (d-i). d-i) Real space conductance map data recorded at $T=6.7$~K at energies $eU_\mathrm{bias}=\pm12$~meV, $\pm8$~meV, and $\pm2.67$~meV. All conductance map data are taken in the same area as shown in (b). The Fourier transformation of the real space conductance map data is shown in the corresponding insets.
}
\label{Fig_2}
\end{figure*}

In spite of all this, LiFeAs has a relatively large critical temperature $T_c$ of about 18~K \cite{Pitcher2010}, supporting the idea that an alternative intrinsic mechanism which enhances the superconducting pairing could be relevant in LiFeAs. 
This material thus is, among the IBS, an ideal candidate to search for evidence of small-momentum nematic mode bosons which couple to the electronic states and stabilize the pairing.
Indeed, small-momentum electronic states in connection to the small three-dimensional FS droplets arising from  hole-like bands (labeled $\alpha$) along the $\Gamma-Z$ direction have been assigned an important role for the superconducting state \cite{Wang2013,Ahn2014a} (Fig.~\ref{FS_schem}(d)).
Interestingly, the superconducting gap has been observed 
to be significantly larger for these $\alpha$-states ($\Delta_1 \approx 6$~meV) as compared to that of the $\beta$- and $\gamma$-bands ($\Delta_2\lesssim4$~meV) \cite{Borisenko2012} which indeed suggests that the strongest pairing interaction in LiFeAs primarily involves states of these bands (c.f. Fig.~\ref{Fig_2}(a) for the signatures of $\Delta_1$ and $\Delta_2$ in low-temperature tunneling spectra).

Previous FT-STS studies of LiFeAs \cite{Haenke2012,Allan2012,Hess2013,Chi2014,Allan2015} have not specifically addressed these $\alpha$-bands in the required energy and temperature range to reveal 
% possible bosonic modes and 
their connection to superconductivity. A possible reason is that these states are located at very small in-plane momenta which requires a particularly high resolution in momentum space. 
In order to achieve this high resolution in our QPI-experiments,
% extract the details of the electronic structure for both occupied and unoccupied states of the $\alpha$-bands we used FT-STS for probing the
% QPI of LiFeAs. Thereby,  we employ a particularly high momentum resolution in order to specifically account for the small in-plane momenta of the states of the $\alpha$-bands near the Fermi level. To this end, 
we recorded large ($110~\text{nm}\times110~\text{nm}$) spectroscopic maps of LiFeAs, where we measured the differential conductance $\mathrm{d}I/\mathrm{d}U(U_\mathrm{bias})$ 
as a direct access to the LDOS (see Appendix~\ref{Sec_Methods} for experimental details). The measurements of these maps have furthermore been performed at several temperatures ranging from 6.7~K up to 25~K to cover the QPI-evolution from the superconducting state to the normal conducting state and to explore a possible temperature evolution of bosonic mode signatures.

\section{Results}

Here we show our experimental results of spectroscopic tunneling measurements of LiFeAs in the superconducting and normal conducting state. From these data we reveal a resonance in the QPI which we assign to a boson representing the nematic fluctuations in LiFeAs.

\subsection{Quasiparticle interference}

Fig.~\ref{Fig_2}(b) depicts representative topographic data of a cleaved surface where the spectroscopic maps have been recorded. In these data one can recognize primarily the typical dumbbell-like iron site defects/impurities (135 defects in total, corresponding to a defect concentration of less than 0.1\% with respect to Fe), which have frequently been observed in LiFeAs \cite{Allan2012,Haenke2012,Grothe2012,Schlegel2017}. 
They serve as the main scattering centers in the sample. 
In Figs.~\ref{Fig_2}(d-i), we present the spectroscopic map data in the superconducting state at 6.7~K for several selected energy values (see Appendix~\ref{Additional_exp} and the Supplementary Material  \cite{SM} for a comprehensive representation of the whole data set).
From these data not only the profound impact of the impurities on the LDOS in their vicinity of several nanometers is apparent. It is also very evident that this impact is strongly energy dependent: 
% The $\mathrm{d}I/\mathrm{d}U$ variation in a given energy slice reveals clear wave-like structures around impurities. 
The relative conductance change around the impurities remains relatively subtle at $E\leq8$~meV. However, at $E=12$~meV, the $\mathrm{d}I/\mathrm{d}U$ variation around the impurities becomes very strong, and acquires a much larger extension.
This is also recognizable in the Fourier transformed data (see Appendix~\ref{Sec_Methods} for a description of the method) shown in the insets of Fig.~\ref{Fig_2}(d-i) (see also Appendix~\ref{Additional_exp}). 
For $E\leq8$~meV, the QPI signal always remains below about 10~pS but significantly exceeds this value at $E=12$~meV.

\begin{figure*}[t]

\centering           

\includegraphics[width=\textwidth]{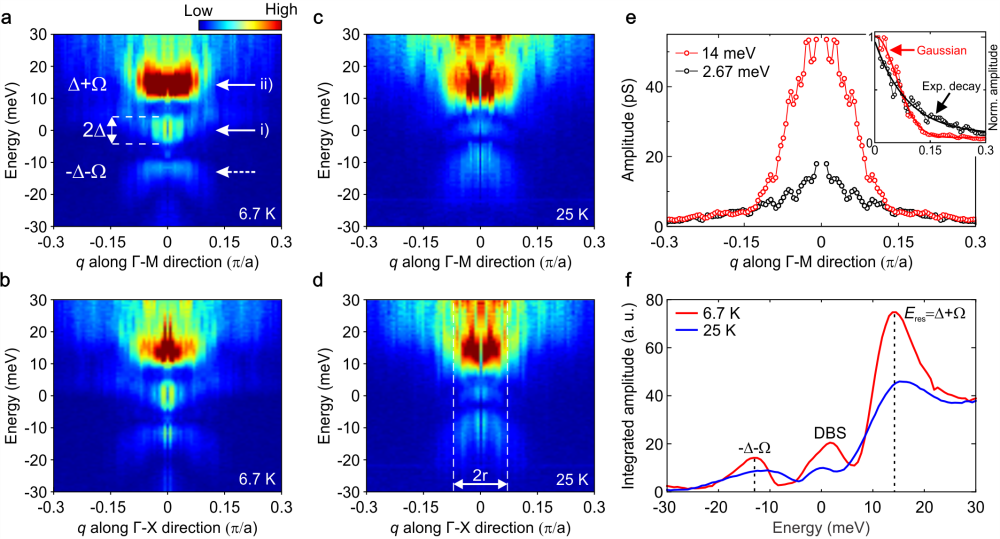}

\caption{Energy and momentum dependence of the FT-STS data in the superconducting and the normal conducting state. a-d) QPI line cuts along the high symmetry directions (c.f. Fig.~\ref{Fig_2}(c)) for the conductance maps taken in the superconducting phase at $T=6.7$~K (a and b) and in the normal phase at $T=25$~K (c and d).
Solid line arrows in (a) indicate the energetic position of features i) and ii), whereas the dashed line arrow indicates the 'replica' of feature ii) at negative energy. e) QPI intensity as a function of $q$ along $\Gamma-M$ of feature ii) and feature i) at 14~meV and 2.67~meV, respectively. Inset: The same data normalized at $q\sim0$. The solid lines represent fits according to an exponential decay $\propto\exp(-\alpha q)$ and a Gaussian for feature i) and feature ii), respectively. 
f) Integrated QPI density at $|q|<r=0.07\pi/a$ (as indicated in (d)) for the 6.7~K and 25~K data. Signatures of the enhanced QPI signal is visible at both positive and negative energy, where a pronounced enhancement occurs in the superconducting state. The FT-STS signature (feature i)) caused by defect bound states (DBS) is visible as well.
\label{Fig_3}
}

\end{figure*}

In order to highlight the pronounced energy and momentum dependence of the Fourier transformed data we plot in Fig.~\ref{Fig_3}(a,b) line cuts of the QPI pattern along the two high symmetric directions as illustrated by the green lines in Fig.~\ref{Fig_2}(c). This data representation reveals a relatively weak anisotropy and two main features [labeled i) and ii) in Fig.~\ref{Fig_3}(a)] at small momenta $q\lesssim0.1\frac{\pi}{a}$ which correspond to  large wavelength modulations in real space. 

Feature i) occurs at $|E|\lesssim 6$~meV, i.e., its energy range coincides with the large superconducting gap $\Delta_1$. This strong in-gap intensity is incompatible with conventional QPI arising from quasiparticle intra- or inter-band scattering processes, obviously because of the absence of quasiparticle states in this energy range. Instead, this structure can straightforwardly be attributed to defect/impurity bound states of LiFeAs. The QPI intensity that arises from these particular states is expected to occur strictly within the superconducting gap energy range. Furthermore,
feature i) decays rapidly with increasing $q$ and thus reproduces other studies which show clearly that the intensity of the bound states emanate outward from the center of an impurity on a length scale of a few nanometers \cite{Grothe2012,Chi2017,Schlegel2017}.
% All these expectations match perfectly with our Fourier transformed data. 
The connection of feature i) to impurity bound states and thus to the superconducting state can be further corroborated by an investigation of its temperature dependence across the critical temperature. Here we observe the feature to fade out, as expected (see Figs.~\ref{Fig_3}(c) and \ref{Fig_3}(d) for $T=25$~K, as well as Appendix \ref{Additional_exp} for intermediate temperature data). At the highest temperature studied ($T=25$~K) there remains just a weak intensity around $E=0$ which accounts for the impact of the impurity on the LDOS in the normal conducting state.

After having established the rather conventional nature of the in-gap intensity, we turn now to analyzing feature ii) which by far dominates the data.
This structure has a sharp onset at about 10~meV and extends up to about 22~meV with a maximum at $E_\mathrm{res}\approx14$~meV. 
It is sharply peaked at $q\sim0$, and has a much larger amplitude than feature i), see Fig.~\ref{Fig_3}(e). Moreover, while the latter decays exponentially in $q$, the momentum dependence of feature ii) is well described by a Gaussian (inset of Fig.~\ref{Fig_3}(e)). This functional variation is remarkable and excludes long wavelength spatial noise as the origin of our observations.
At first glance, the occurrence of such a strong intensity in this energy range significantly away from the superconducting gap appears surprising. Clearly, it cannot be the direct signature of an impurity bound state because such a state would exist strictly only within the superconducting gap energy \cite{Balatsky2006}. Furthermore, conventional QPI arising from intra- or inter-band scattering processes within the two $\alpha$-bands (which in principle would be compatible with the relatively small $q$-value) at first glance cannot account for this observation: It is well known that the $\alpha$-bands possess a strong $k_z$ dispersion \cite{Wang2013,Borisenko2016}. Thus, the QPI signal that emerges from these bands normally should be very broad and featureless in energy and momentum.
This is because the QPI that is measured by an STM is \textit{a priori} not sensitive to the $k_z$. At a given energy, the measured QPI pattern is expected to be a result of the superposition of all different in-plane (i.e. $\Delta k_z=0$) scattering wave vectors at different $k_z$ and the in-plane projection of scattering vectors with finite $\Delta k_z$. Since the scattering vectors must combine points of equal energy, a $k_z$-dispersion is always related to a certain broadening in the $(k_x,k_y)$-direction.
Indeed, one might conjecture that the faint and broad QPI structures of rather low amplitude (lower than $\sim5$~pS in Fig.~\ref{Fig_2},
see, e.g. panel (h) of the figure) are compatible with this picture. In contrast,
the observed extraordinary enhancement of intensity in the particular energy range of feature ii) cannot be explained in this way and therefore directly implies an unusual amplification of the measured QPI.
In fact, feature ii) can straightforwardly be interpreted as resonantly enhanced QPI due to a boson-assisted renormalization of scattering potential as is illustrated in Figs.~\ref{FS_schem}(a) and \ref{FS_schem}(b).
More specifically, this scenario implies feature ii) to be caused by bosons centered at energy $\Omega$ and momentum $q\sim0$, where $E_\mathrm{res}=\Delta+\Omega$, with $\Delta$ the superconducting gap 
(Appendix~\ref{Subsec_theory}).

Before we discuss the possible implications of the detection of pertinent bosonic excitations with small momentum, we investigate further corroborations of this fundamental finding. Firstly, the electron-boson interaction must concern not only the unoccupied states (as discussed so far) but also the occupied electronic states. Indeed, the close inspection of the data shown in Figs.~\ref{Fig_3}(a) and \ref{Fig_3}(b), reveals, despite an overall weaker amplitude, a pronounced enhancement of the QPI signal at about $-E_\mathrm{res}=-\Delta-\Omega$ (dashed arrow).
Secondly, feature ii) is significantly more intense than robust QPI signatures at larger wave vectors. This can be inferred from the additional data set in Appendix~\ref{Standard_QPI}, where we explicitly compare the integrated intensity of feature ii) with well known nested intraband scattering within the $\gamma$ band.

Interestingly, the signature of the boson persists in the whole temperature regime up to the normal conducting state far above the critical temperature at $T=25$~K, as is revealed by the inspection of our temperature dependent QPI data, see Figs.~\ref{Fig_3}(c), \ref{Fig_3}(d), and also Fig.~\ref{MDC2} in Appendix~\ref{Temp_data}.
The difference between the superconducting and normal conducting states is further illustrated in Fig.~\ref{Fig_3}(f), where we show the integrated QPI intensity over the region with $|q|<r=0.07\frac{\pi}{a}$ (c.f. Fig.~\ref{Fig_3}(d)) as a function of  the energy for both $T=6.7$~K and $T=25$~K.
The data for the superconducting state show clearly that there are resonance-like peaks with a quite sharp onset at around $\pm10$~meV with peak values at about $\pm(13\dots14)$~meV, where the peak at positive energy is much more pronounced. Both peaks broaden in the normal conducting state where the low-energy edge shifts to about $\pm5$~meV, whereas the peak positions, in particular that of the better resolved peak at positive energy barely change. 
Qualitatively, the sharpening of the  peaks in the superconducting state can be rationalized as a direct consequence of the formation of Bogoliubov quasiparticle states and a further, resonance-like enhancement of the QPI signal at $E_\mathrm{res}=\Delta+\Omega$ by an exaggeration of the scattering potential due to coupling to the bosons. 
We extract, by focusing on the better resolved peak in the superconducting state at positive energy where $E_\mathrm{res}=14\pm4$~meV and by employing the gap at the $\alpha$ states $\Delta_1=6$~meV \cite{Borisenko2012}, a mode peak energy $\Omega=8\pm4$~meV \cite{error}.

At first glance it seems surprising that the resonance peak remains practically unshifted in energy upon entering the normal state. In order to obtain further insight into the nature of our observation we performed a careful theoretical analysis of the impurity scattering in LiFeAs using realistic parameters for band structure (including the spin-orbit coupling \cite{Borisenko2016}), scattering potential, and electron-boson coupling (see Appendix~\ref{Subsec_theory_appl}). A central finding of this analysis is that, due to the particular structure of the $\alpha$-bands in LiFeAs, which encompasses a spin-orbit coupling induced separation of the band maxima with high density of states by approximately the same amount (of the order of 10~meV) than $\Delta_1$, the resonance conditions in the normal state are accidently similar to that of the superconducting state. Indeed, the analysis yields a consistent description of the observed resonance in both phases, explaining the absence of a shift of the peaks. A further important and interesting result of this analysis is that the considered boson in interaction with the band structure of LiFeAs leads to a stable superconducting solution with the leading experimental gap value at the $\alpha$-states.

It further is interesting to verify our finding of a small-momentum boson against optical spectroscopy where signatures of electron-boson coupling at $q\sim0$ should be well detectable. Indeed, a recent optical study \cite{Hwang2015} which reveals a $q=0$ mode at the very same energy underpins our finding, however, lacking the general momentum sensitivity and resolution which is provided by the analysis of resonantly enhanced QPI as presented in this work.

\subsection{Comparison with tunneling spectroscopy}

\begin{figure}[h]

\centering           

\includegraphics[width=\columnwidth]{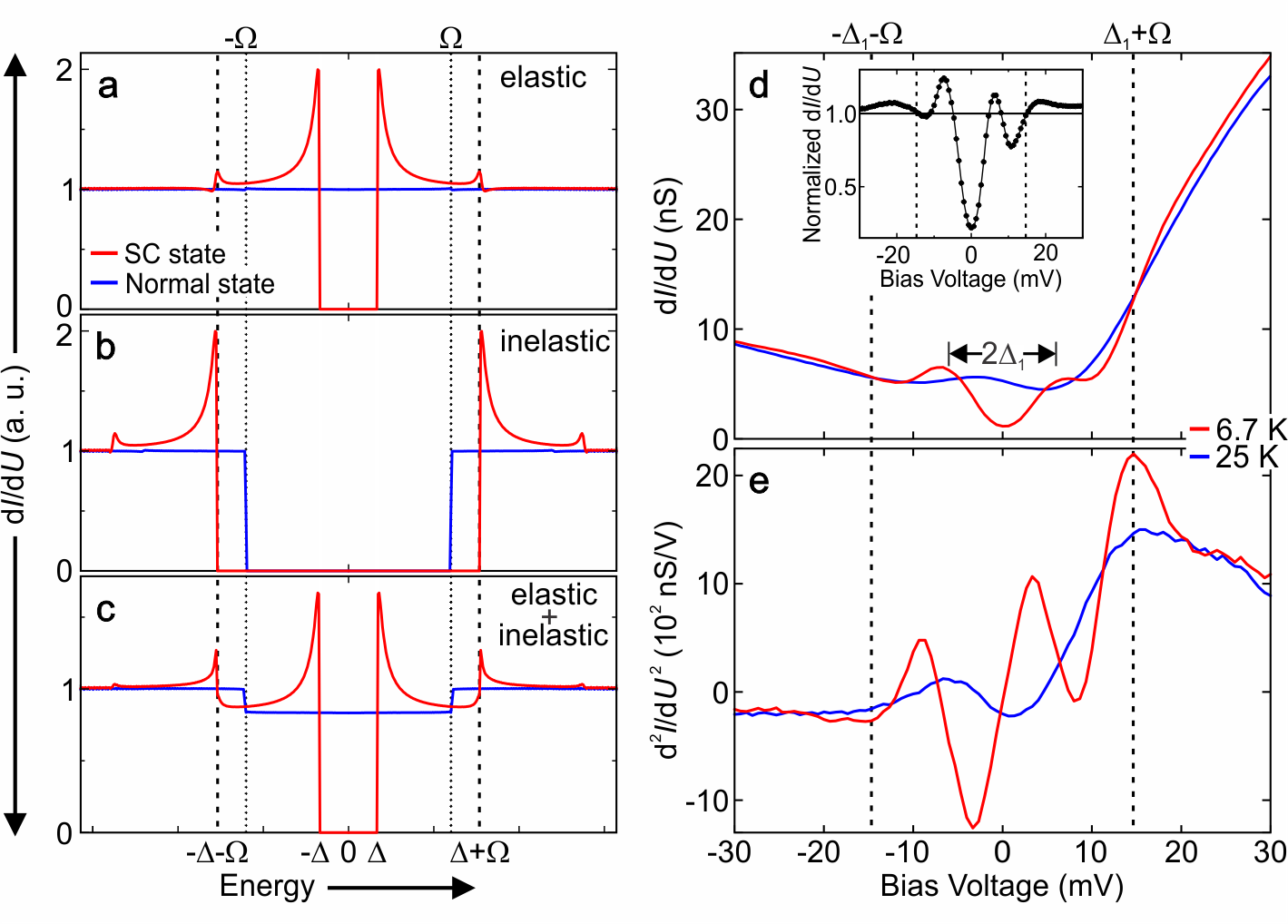}

\caption{Impact of bosonic excitations on tunneling spectra. a-c) Calculated elastic (a) and inelastic (b) contributions to the $\mathrm{d}I/\mathrm{d}U$ spectrum both in the superconducting (SC) and the normal conducting state (red and blue lines, respectively) for the simplified model of electron-boson coupling (boson energy $\Omega$, superconducting gap $\Delta$) as described in Appendix~\ref{Subsec_theory}. The theoretical approach to the inelastic contributions is taken from Ref.~\onlinecite{Hlobil2017}. For the total tunneling spectrum shown in (c) the ratio of elastic and inelastic contribution is fixed by the chosen model parameters. The combination of the two different contributions leads in an energy range between $\Omega$ and $\Delta + \Omega$ to the characteristic depletion of spectral weight in the superconducting state with respect to the normal conducting state. This depletion results in a characteristic dip-hump structure in normalized tunneling spectra \cite{Hlobil2017}. d) Average $\mathrm{d}I/\mathrm{d}U$ spectra measured on defect free surface areas in the superconducting (6.7~K, red) and normal conducting state (25~K, blue). Inset: $\mathrm{d}I/\mathrm{d}U$ spectrum at 6.7~K normalized with respect to the normal conducting state spectrum at 25~K revealing the dip-hump structure. e) Voltage derivative $\mathrm{d}^2I/\mathrm{d}U^2$ of the data in (d). The dashed lines in both polarities through (a) to (e) indicate the inelastic peak position.
}
\label{Fig_4}
\end{figure}

Another, alternative way to identify the signature of bosonic excitations is the investigation of tunneling $\mathrm{d}I/\mathrm{d}U$ spectra far away from impurities, where
bosonic excitations may leave their fingerprints in two fundamentally different ways:
On the one hand, bosonic excitations which couple to Bogoliubov quasiparticles of a superconductor may induce a characteristic fingerprint  in the tunneling spectra if the coupling is strong enough. More specifically, a well-defined bosonic mode at energy $\Omega$ is expected to give rise to a peak structure at the energy $\Delta +\Omega$ (with $\Delta$ the superconducting gap) \cite{McMillan1965,Scalapino1966}, see Fig.~\ref{Fig_4}(a). 
On the other hand, bosonic excitations may open up a relevant \textit{inelastic} scattering channel in addition to the usual elastic one, playing a dominant role in the $\mathrm{d}I/\mathrm{d}U$ in unconventional superconductors, and particularly in LiFeAs \cite{Hlobil2017}. In the case of a relevant inelastic tunneling contribution due to a well defined boson, a significant enhancement of the tunneling $\mathrm{d}I/\mathrm{d}U$ is expected for $E>\Delta+\Omega$ and $E>\Omega$ in the superconducting and the normal conducting states, respectively (Fig.~\ref{Fig_4}(b)). As is shown in Figs.~\ref{Fig_4}(b) and \ref{Fig_4}(c), this leads to a characteristic depletion and step-like enhancement of the $\mathrm{d}I/\mathrm{d}U$ in the superconducting state with respect to that of the normal conducting state at $E<\Delta+\Omega$ and $E>\Delta+\Omega$, respectively. 

We therefore show in Fig.~\ref{Fig_4}(d) a direct comparison of the $\mathrm{d}I/\mathrm{d}U$ for both the superconducting state and the normal conducting state and investigate the data for fingerprints of the boson that we have identified from the QPI data. 
Quite clearly, the data show dominant signatures of inelastic tunneling: At energies close to the gap edges $\mathrm{d}I/\mathrm{d}U$ of the superconducting state exhibits a depletion with respect to that of the normal state whereas it shows a step-like enhancement at about 14~meV and exceeds the $\mathrm{d}I/\mathrm{d}U$ of the normal state beyond. This energy dependence leads to the known characteristic 'dip-hump'  anomaly in normalized $\mathrm{d}I/\mathrm{d}U$ data \cite{Hlobil2017}, which is often observed in various unconventional superconductors, including LiFeAs  \cite{Hudson1999,Jenkins2009,Wang2013d,Song2014,Chi2012,Nag2016,Chi2017} (inset of Fig.~\ref{Fig_4}(d)).
Since the step-like increase of the $\mathrm{d}I/\mathrm{d}U$ in the superconducting state at about 14~meV (c.f. Fig.~\ref{Fig_4}(e)) is expected to occur at $\Delta+\Omega$ we  extract $\Omega=8\pm4$~meV if we use $\Delta_1=6$~meV as the leading gap \cite{error}.  Thus, this completely different approach of accessing the bosonic excitations in LiFeAs yields a boson energy in excellent agreement with our QPI analysis.
This suggests that the salient above-gap structure in the local $\mathrm{d}I/\mathrm{d}U$ tunneling spectra of LiFeAs result from the \textit{same} small-wavevector bosonic mode which we infer from our QPI data. Scenarios which interpret the nature of the dip-hump structure in terms of an antiferromagnetic spin resonance \cite{Chi2012,Hlobil2017,Chi2017} can therefore be excluded. 
It is interesting to point out that the intrinsic width of the step is not significantly reduced in tunneling spectra at very low temperature (see Appendix~\ref{Additional_exp} for data at 300~mK) despite a significant sharpening of the thermal width of the coherence peaks. This suggests that the energetic width of the involved boson is not sharp, indicative of the importance of many-body effects for the nature of the boson.

\section{Conclusions}

Our identification of bosonic modes at $\Omega\approx8$~meV with a small wave vector $q$ and a connected resonance-like enhancement of the QPI signal in the superconducting state provides fresh input and constraints for rationalizing the pairing mechanism of LiFeAs \cite{Brydon2011,Wang2013,Ahn2014a,Yin2014,Saito2015}. The observed incommensurate spin fluctuations definitely can be excluded as a microscopic origin for our observations because inelastic neutron scattering proves a too large wave vector and a 
negligible difference between the normal and the superconducting states \cite{Qureshi2012}.  A further alternative but rather exotic origin of our bosonic mode could be small-$q$ spin-fluctuations, which have been derived in model calculations \cite{Brydon2011}, but have not yet experimentally been observed. The final remaining microscopic  origin of the mode which is consistent with our analysis lies in 
\textit{dynamic} nematic fluctuations, i.e., small-$q$ electronic density fluctuations the ordered phase of which has been observed experimentally in the context of static density fluctuations with broken rotational symmetry (stripes) \cite{Li2017,Yim2018}.

In this context it is important to emphasize recent STM results of Ref.~\onlinecite{Yim2018} on strained LiFeAs, where a static form of the nematic fluctuations, i.e., a rotational symmetry broken phase with long-range small-wavevector density variations, and a concomitant suppression of superconductivity is observed.
The interpretation of small-momentum excitations observed in our study as dynamic nematic fluctuations is therefore strongly corroborated. Furthermore, the reported suppression of superconductivity upon the onset of charge density wave order supports our conclusion that the ``nematic'' small-wavevector density fluctuations are crucial for sustaining superconductivity in LiFeAs.

% The boson-assisted resonant enhancement of the QPI amplitude presented in this work is a very general effect. In principle, it is not limited to bosons with a particular $q$ value, and it should be applicable to investigate the electron-boson coupling in any material suitable for FT-STS and therefore constitutes a new method with an applicability clearly beyond the present study.
% In this vein, 
% it seems on the one hand worthwhile to further scrutinize the FT-STS of LiFeAs and to search for signatures of boson-enhanced QPI amplitudes in the whole Brillouin zone. This includes the reinvestigation of subtle geometric band renormalizations close to the antiferromagnetic wave vector \cite{Allan2015}. On the other hand, 
% In view of the peculiar position of LiFeAs among the IBS, our finding of a strong small-$q$ bosonic mode poses the question whether such a mode is also detectable in other IBS, in particular those where the evidence for FS nesting and an antiferromagnetic resonance is strong, e.g. in doped BaFe$_2$As$_2$ and NaFeAs.
% The perspective, to thus obtain a comprehensive account of the bosonic excitations which couple to the different electronic bands appears a promising route for providing the experimental data basis necessary for testing model calculations (based on orbital and spin fluctuations) for the superconducting mechanism of LiFeAs, and the IBS in general \cite{Wang2013,Ahn2014a,Saito2015}.

\section*{Acknowledgements}

We acknowledge fruitful discussions with S. Borisenko, M. Braden, I. Eremin, D. van der Marel, M. Vojta, and P. Wahl. We further thank U. Nitzsche, U. Gr\"afe, and D. Baumann for technical assistance.
  This project has been supported by the Deutsche Forschungsgemeinschaft through the Priority Programme SPP1458 (Grant HE3439/11 and BU887/15-1) and through GRK 1621.  S.W. acknowledges funding by DFG under the Emmy-Noether program (Grant No. WU595/3-3). T.H. acknowledges support by the DFG under Grant No. HA 6037/2-1. Furthermore, this project has received funding from the European Research Council (ERC) under the European Unions' Horizon 2020 research and innovation programme (grant agreement No 647276 -- MARS -- ERC-2014-CoG). 

\begin{appendix}

\section{Theory}

Here we present our theoretical approach to the observed resonance feature in the QPI. Starting from a general treatment of impurity scattering in the presence of electron-boson interaction we show that an exaggerated effect is particularly obtained in the specific situation of LiFeAs.

\subsection{Renormalization of the scattering potential}

\label{Subsec_theory}

We consider one single local impurity embedded in a system of conduction electrons which
additionally couples to a system of bosons. Of particular interest is the change of the electronic
local density of states in the environment of the impurity due to elastic scattering of quasiparticles
interacting via virtual bosonic excitations. The calculated effect to the local density of states
variations is compared with STM spectroscopy measurements in LiFeAs. To avoid taking into
account additional effects from the band structure we consider here a single parabolic band and
the simplest possible form of electron-boson coupling and impurity scattering. The model Hamiltonian for such a system consists of three parts, $\mathcal{H} = \mathcal{H}_0 + \mathcal{H}_{V} + \mathcal{H}_{eb}$, where
\begin{eqnarray*}
\mathcal{H}_0 &=& \sum_{{\bf k},{\sigma}} \varepsilon_{\bf k} c_{{\bf k},\sigma}^\dagger c_{{\bf k},\sigma}^{} + \sum_{\bf q} \omega_{\bf q} b_{\bf q}^\dagger b_{\bf q}^{} , \\
\mathcal{H}_{eb} &=&  \frac{1}{\sqrt{N}}\sum_{{\bf k},{\bf q},\sigma} g_{{\bf k},{\bf q}} \left( b_{\bf q}^\dagger c_{{\bf k},\sigma}^\dagger c_{{\bf k}+{\bf q},\sigma}^{} + b_{\bf q} c_{{\bf k}+{\bf q},\sigma}^{\dagger} c_{{\bf k},\sigma} \right),  \\
\mathcal{H}_V &=& \frac{1}{N}\sum_{{\bf k},{\bf q},\sigma} V_{{\bf k},{\bf q}} \left(  c_{{\bf k},\sigma}^\dagger c_{{\bf k}+{\bf q},\sigma}^{} + c_{{\bf k}+{\bf q},\sigma}^{\dagger} c_{{\bf k},\sigma} \right).
\end{eqnarray*}
Here, $N$ is the number of lattice sites. The first term describes a system of free conduction electrons and bosons. Thereby, the operator $c_{{\bf k},\sigma}^\dagger$ creates an electron with momentum ${\bf k}$ (dispersion $\varepsilon_{\bf k}$) and spin $\sigma$ and the operator $b_{\bf q}^\dagger$ creates a boson with momentum ${\bf q}$ (dispersion $\omega_{\bf q}$). The coupling between electrons and bosons is represented by the second term $\mathcal{H}_{eb}$. It describes the scattering of an electron between states ${\bf k}$ and ${\bf k} + {\bf q}$ while a boson with momentum ${\bf q}$ is created or annihilated. The corresponding parameter of the coupling strength $g_{{\bf k},{\bf q}}$ generally depends on the 
contributing momentum vectors accounting for 
a possible non-local electron-boson interaction. The term $\mathcal{H}_{V}$ describes the scattering interaction off the single impurity with the momentum-dependent scattering potential $V_{{\bf k},{\bf q}}$. Note that this part breaks the translation symmetry of the Hamiltonian.

The variations of the local density of states due to the impurity scattering in an electron-boson coupled system is calculated as follows. At first the Hamiltonian $\mathcal{H} = \mathcal{H}_0 + \mathcal{H}_{V} + \mathcal{H}_{eb}$ is mapped to a particular effective Hamiltonian $\tilde{\mathcal{H}} = \tilde{\mathcal{H}_0} + \tilde{\mathcal{H}_{V}}$ which is constructed in such a way that the electron-boson coupling is fully integrated out by use of a unitary transformation. The new Hamiltonian has the same form as the original one,
\begin{eqnarray}
\tilde{\mathcal{H}} &=& \sum_{{\bf k},{\sigma}} \tilde{\varepsilon}_{\bf k} c_{{\bf k},\sigma}^\dagger c_{{\bf k},\sigma}^{} + \sum_{\bf q} \tilde{\omega}_{\bf q} b_{\bf q}^\dagger b_{\bf q}^{} \nonumber \\ 
&+& \frac{1}{N}\sum_{{\bf k},{\bf q},\sigma} \tilde{V}_{{\bf k},{\bf q}} \left(  c_{{\bf k},\sigma}^\dagger c_{{\bf k}+{\bf q},\sigma}^{} + c_{{\bf k}+{\bf q},\sigma}^{\dagger} c_{{\bf k},\sigma} \right)
\label{Heff}
\end{eqnarray}
but with renormalized energy parameters $\tilde{\varepsilon}_{\bf k}$, $\tilde{\omega}_{\bf q}$, and $\tilde{V}_{{\bf k},{\bf q}}$. It is calculated by using the Projective Renormalization Method (PRM) \cite{Becker2002} which has already been successfully applied to solve models with electron-boson interaction \cite{Sykora2005,Cho2016}. Note that the form of Hamiltonian \eqref{Heff} is strictly only valid in the normal conducting state. For ordered states the inclusion of symmetry breaking order parameter terms is necessary. \cite{Cho2016} Here we focus on the renormalization of the impurity potential in the third term of Eq.~\eqref{Heff} which can be discussed most clearly in the normal conducting state. The influence of the superconducting order is discussed further below.

Within the PRM approach non-linear difference equations for the renormalized parameters (renormalization equations) are numerically evaluated starting from the given energy parameters of the original Hamiltonian $\mathcal{H}$. The effective Hamiltonian \eqref{Heff} is then taken to calculate the local density of states variations using the standard t-matrix method.

Main effect of the electron-boson coupling is the renormalization of the impurity scattering potential.  This can be seen from the result of the renormalized impurity potential $\tilde{V}_{{\bf k},{\bf q}}$ in {\it lowest order perturbation theory} with respect to the original coupling parameters $g_{{\bf k},{\bf q}}$ and $V_{{\bf k},{\bf q}}$. The perturbation theory can be easily carried out within the PRM by following the ideas of Ref.~\onlinecite{Becker2002}. Considering for simplicity momentum-independent coupling parameters of the original Hamiltonian, $g_{{\bf k},{\bf q}} = g$ and $V_{{\bf k},{\bf q}} = V$,  the lowest order result for the renormalized impurity potential can be written in the form $V_{{\bf k},{\bf k}'}^{(2)} = (\tilde{V}_{{\bf k},{\bf k}'}^{(2)} + \tilde{V}_{{\bf k}',{\bf k}}^{(2)})/2$ where 
\begin{eqnarray}
\tilde{V}_{{\bf k},{\bf k}'}^{(2)} &=& V + \frac{Vg}{N}\sum_{\bf q} \left[ \frac{2f_{{\bf k}'+{\bf q}} - 1}{\varepsilon_{{\bf k}'+{\bf q}}-\varepsilon_{{\bf k}'}+\omega_{\bf q}} \times \right. \nonumber \\ 
&\times& \left(\frac{g}{\varepsilon_{{\bf k}'+{\bf q}}-\varepsilon_{{\bf k}'}+\omega_{\bf q}} 
- \frac{g}{\varepsilon_{{\bf k}+{\bf q}}-\varepsilon_{{\bf k}}+\omega_{\bf q}}\right) \nonumber \\ \nonumber \\
&+& \frac{2f_{{\bf k}+{\bf q}} - 1}{\varepsilon_{\bf k} - \varepsilon_{{\bf k}+{\bf q}}+\omega_{\bf q}} \times \nonumber \\ 
&\times& \left. \left(\frac{g}{\varepsilon_{{\bf k}'} - \varepsilon_{{\bf k}'+{\bf q}}+\omega_{\bf q}} - \frac{g}{\varepsilon_{{\bf k}} -\varepsilon_{{\bf k}+{\bf q}}+\omega_{\bf q}}\right)\right].
\label{V_pert}
\end{eqnarray}
The function $f_{\bf k} = 1/(1 + e^{\beta (\varepsilon_{\bf k} - \varepsilon_{F})})$ denotes the Fermi distribution with inverse temperature $\beta = 1 / (k_B T)$ and Fermi energy $\varepsilon_{F}$. Terms proportional to the boson distribution function also arise in the second order perturbation theory but they  can be neglected at low temperatures due to very small boson occupation. The expression \eqref{V_pert} diverges below a characteristic temperature for particular values of the momentum vectors ${\bf k}$ and ${\bf k}'$. The divergence appears since at certain momentum vectors ${\bf q}$ in the summation the energy denominators become zero while at the same time the Fermi factors $(2f_{{\bf k}'+{\bf q}} - 1)$ and $(2f_{{\bf k}+{\bf q}} - 1)$ change their sign.

Since this behavior is essential for the observed resonance-like enhancement of the tunneling density of states we here explain this renormalization process in more detail. Let us simplify the discussion by considering dispersionless bosons, i.~e.~$\omega_{\bf q} = \Omega$. Furthermore, we consider the renormalized scattering potential in Eq.~\eqref{V_pert} at a particular  fixed momentum vector ${\bf k}'$ such that $\varepsilon_{{\bf k}'} = \varepsilon_F + \Omega$. In this case the denominator in the first line of Eq.~\eqref{V_pert} becomes $\varepsilon_{{\bf k}'+{\bf q}}-\varepsilon_{{\bf k}'}+\omega_{\bf q} = \varepsilon_{{\bf k}'+{\bf q}} - \varepsilon_{F}$. Thus, during the summation over ${\bf q}$ this denominator becomes zero for $\varepsilon_{{\bf k}'+{\bf q}} = \varepsilon_{F}$ but changes its sign which usually leads to a cancellation of diverging terms. Here, however, the situation is different. Due to the presence of the Fermi distribution the numerator in the first line, $2f_{{\bf k}'+{\bf q}} - 1$, changes its sign exactly at the same ${\bf q}$ namely for $\varepsilon_{{\bf k}'+{\bf q}} = \varepsilon_{F}$. Since the factor in brackets in the second line does not change its sign at this particular ${\bf q}$ (for ${\bf k} \ne {\bf k}'$) the sign of the diverging terms is preserved and we here have a real divergency which cannot be canceled out by summation.  
From a close inspection of all terms in Eq.~\eqref{V_pert} with respect to the above considerations one can conclude that the renormalized scattering potential becomes strongly enhanced when the momentum vectors fulfill roughly the 'resonance conditions' $\varepsilon_{F}-\varepsilon_{{\bf k}'} = \pm\Omega$ and $\varepsilon_{F}-\varepsilon_{{\bf k}} = \pm\Omega$.

\begin{figure}[t]
\centering           
\includegraphics[width=\columnwidth]{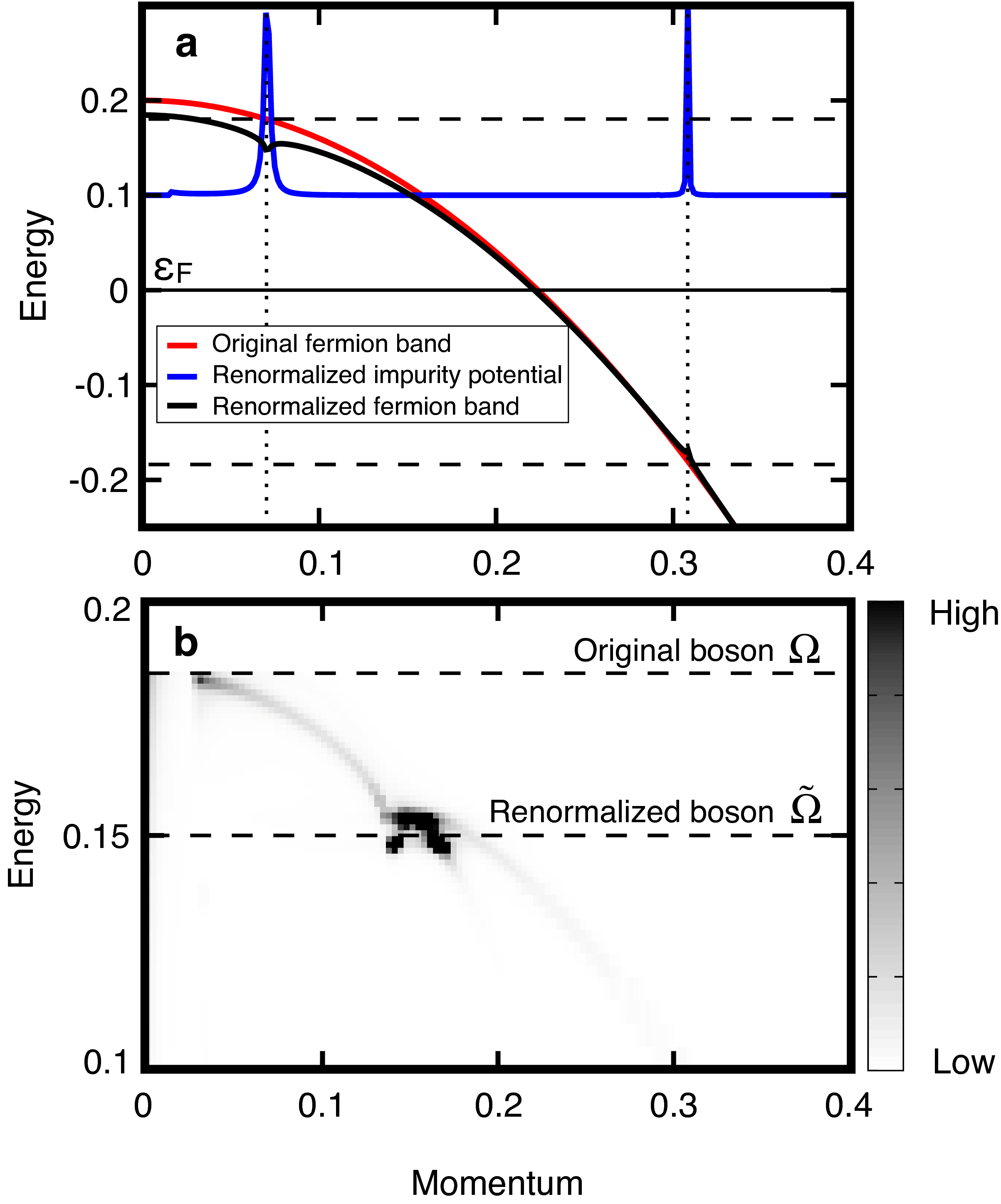}
\caption{Renormalization effects due to electron-boson coupling. a) Renormalized fermion dispersion $\tilde{\varepsilon}_{\bf k}$ (black solid line) and impurity potential $\tilde{V}_{{\bf k},-{\bf k}} $ (blue  solid line) combining opposite momentum vectors representing relevant elastic scattering processes calculated for a simplified hole like dispersion (red solid line). The momentum vector runs along the cut ${\bf k} = (k,0)$. The particular ${\bf k}$ points  which fulfill the resonance condition $\varepsilon_{F}-\varepsilon_{{\bf k}} = \pm\Omega$ (indicated by the intersection points of the dashed lines with the original fermion band) give rise to strong renormalization of the impurity potential as well as the fermionic dispersion (kink-like structures). These pronounced momentum vectors are indicated  by dotted lines. b) Fourier-transformed local density of states as a function of scattering momentum calculated by a t-matrix method using the renormalized energy parameters from Eq. (1). The intensity is strongly enhanced around the particular scattering momentum combining the intersection points shown in panel (a). This exaggeration appears at the renormalized boson energy $\tilde{\Omega}$ which has decreased with respect to the original value $\Omega$ (dashed lines).
}
\label{Fig: Theory}
\end{figure}

Thus, the dominant scattering vectors in the presence of an electron-boson coupling are determined by the intersection points of the electron dispersion $\varepsilon_{\bf k}$ and the boson energy $\omega_{\bf q}$ (see Fig.~\ref{Fig: Theory}). The corresponding Feynman diagram of this process is shown in the inset of Fig.~\ref{FS_schem}(b). Note that in the actual numerical treatment the divergence is removed by taking into account contributions to the renormalization up to infinite order. Further note that the above considerations are also valid for a general momentum-dependent boson energy $\omega_{\bf q}$.

The same processes lead also to a renormalization of the electron dispersion. This can be seen again from its second order perturbation theory result which has a similar form as expression \eqref{V_pert}. As seen from Fig.~\ref{Fig: Theory}(a) at the intersection points between the bare dispersion and the boson energy the renormalization is strongest and gives rise to a kink-like structure. This feature is also well-known from several ARPES studies in cuprate and pnictide materials.

We have calculated all renormalized quantities in Eq.~\eqref{Heff} starting from a simplified model Hamiltonian $\mathcal{H}$ with fixed parameters describing roughly the situation relevant to LiFeAs. The parameters of the free part $\mathcal{H}_0$ (related to some relevant energy unit) are a 2D parabolic hole-like dispersion $\varepsilon_{\bf k} = -k^2+0.2$ and a momentum-independent boson energy $\omega_{\bf q} = \Omega = 0.18$ lying slightly below the top of the fermion band. For the coupling parameters we have chosen the momentum-independent values $V=0.1$ and $g=0.1$. The numerical results are shown in Fig.~\ref{Fig: Theory}(a). As already observed in the perturbation theory discussed above a strong renormalization is found at the particular ${\bf k}$ points where the original fermion band $\varepsilon_{{\bf k}}$ (red solid line) intersects with the values $\pm\Omega$ (dashed lines). This leads to kink-like structures in the renormalized fermion dispersion (black solid line) and, most importantly, to a strong enhancement of the elastic scattering potential (blue solid line). Thus, in the presence of electron-boson interaction the effective impurity scattering potential becomes strongly momentum-dependent for particular scattering momentum which is characterized by the resonant coupling to a virtual bosonic mode 
(compare inset of Fig.~\ref{FS_schem}(b)). 

The calculated renormalized quantities $\tilde{\varepsilon}_{\bf k}$ and $\tilde{V}_{{\bf k},{\bf q}}$ can be used as input parameters for a subsequent standard t-matrix approach to calculate the Fourier-transformed local density of states,
\begin{eqnarray}
\rho ({\bf q},E) = \frac{1}{\pi} \sum_{\bf k} \Im  G({\bf k},{\bf k} - {\bf q},E),
\label{LDOS}
\end{eqnarray}
which is the quantity that is directly measured by STM/STS experiments. $G({\bf k},{\bf k}',E)$ is the retarded Greens function in the presence of one single impurity and is related to the retarded Greens function $G_0({\bf k},E)$ of the bulk material via the equation \cite{Balatsky2006}
\begin{eqnarray}
G({\bf k},{\bf k}',E) = G_0({\bf k},E) + G_0({\bf k},E) T_{{\bf k},{\bf k}'}(E)G_0({\bf k}',E), \nonumber \\
\label{G_imp}
\end{eqnarray}
where the energy-dependent t-matrix $T_{{\bf k},{\bf k}'}(E)$ is determined by the following self-consistency equation, 
\begin{eqnarray}
T_{{\bf k},{\bf k}'}(E) = \tilde{V}_{{\bf k},{\bf k}'} + \sum_{{\bf k}''} \tilde{V}_{{\bf k},{\bf k}''} G_0({\bf k}'',E)T_{{\bf k}'',{\bf k}'}(E).
\label{t-matrix}
\end{eqnarray}
The non-interacting Greens function $G_0({\bf k},E) = (E - \tilde{\varepsilon}_{\bf k} - i\delta)^{-1}$ contains the renormalized fermion dispersion.

Using as input parameters for our theory the two functions $G_0({\bf k},E)$ and $\tilde{V}_{{\bf k},{\bf k}'}$ we have solved the system of Eqs.~\eqref{G_imp} and \eqref{t-matrix} self-consistently. The obtained result for the full Greens function $G({\bf k},{\bf k}',E)$ is inserted in Eq.~\eqref{LDOS} in order to evaluate the intensity $\rho ({\bf q},E)$. The numerical result is shown in Fig.~\ref{Fig: Theory}(b). In a large energy range the intensity is slightly enhanced for the scattering vectors combining momentum vectors with same energy (conventional elastic scattering). There, the value of the scattering potential is nearly constant and therefore the intensity (gray shaded area) is mainly determined by the fermionic density of states while its momentum dependence is given by the dispersion $\tilde{\varepsilon}_{\bf k}$ of the renormalized band. However, at scattering momentum nearly equal to the distance between the two inner peaks of $\tilde{V}_{{\bf k},-{\bf k}}$ in Fig.~\ref{Fig: Theory}(b) the intensity is strongly enhanced (black area) due to the exaggeration of the renormalized scattering potential. Moreover, fine structures are visible which arise from the kink structure of $\tilde{\varepsilon}_{\bf k}$ around the resonance points. Note that similar considerations have been also applied to study the pinning of dynamic spin-density-wave fluctuations where strong modulations in the local density of states could be traced back to the interaction with a correlated background medium \cite{Polkovnikov2002,Rossi2010}.

Energetically, the resonance appears inside the kink feature of $\tilde{\varepsilon}_{\bf k}$ which corresponds also to the minimum value of the renormalized boson energy $\tilde{\Omega} = \mbox{min}(\tilde{\omega}_{\bf q})$. The numerical result of this energy level is shown in Fig.~\ref{Fig: Theory}(b) by the lower dashed line. For the specific parameter values chosen in our calculation the momentum dependence of the  renormalized boson energy $\tilde{\omega}_{\bf q}$ is rather weak and is therefore not shown here. Note, however, that such a dispersion may become important if the system is very close to a transition to ordered states \cite{Sykora2005}.

\begin{figure}[t]
\centering           
\includegraphics[width=\columnwidth]{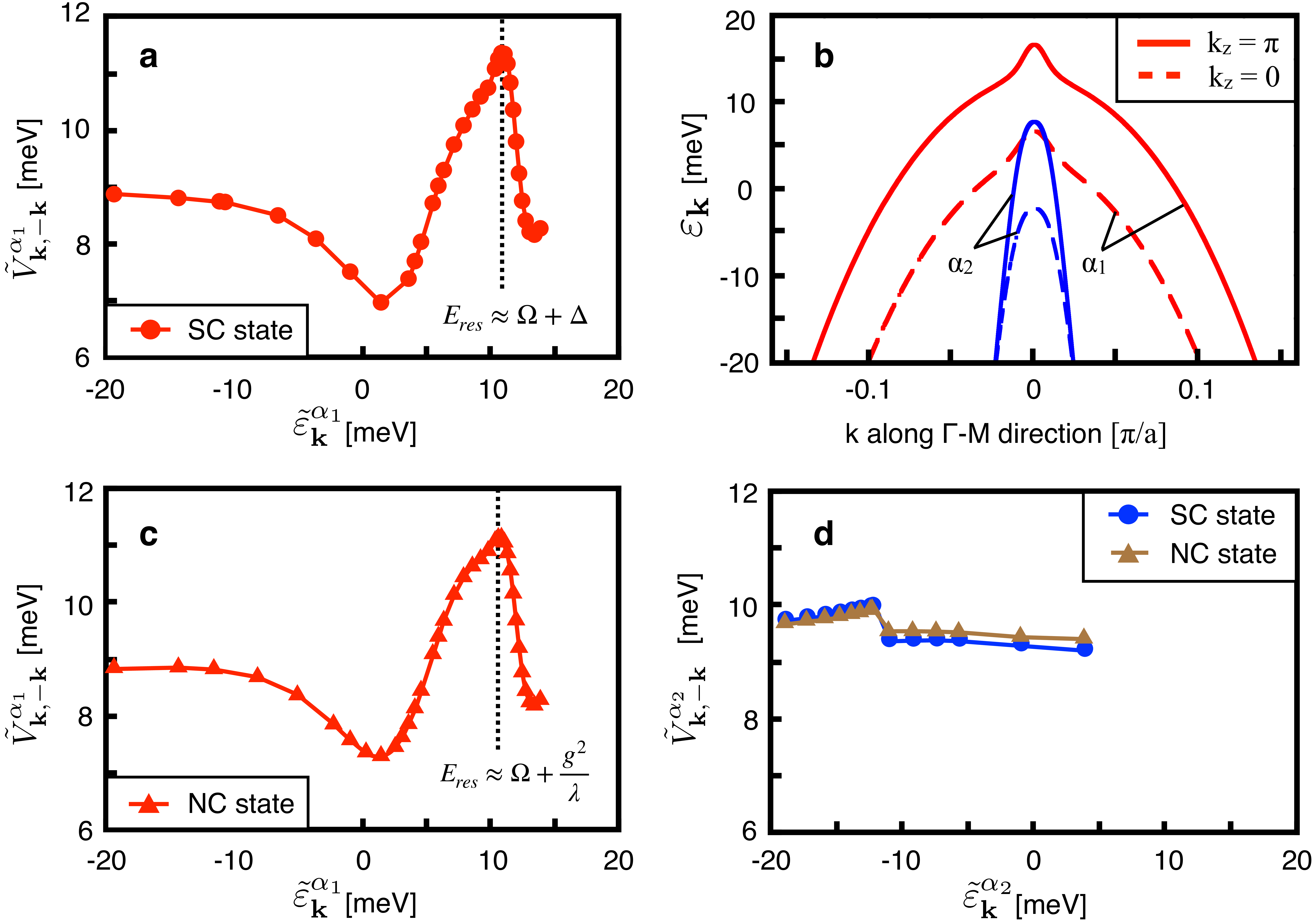}

\caption{\label{Fig: Theory_2}Renormalized scattering potential in LiFeAs. a) Renormalized impurity potential $\tilde{V}_{{\bf k},-{\bf k}}^{\alpha_1}$ combining opposite momentum vectors on the larger hole pocket $\alpha_1$ plotted against the corresponding renormalized dispersion $\tilde{\varepsilon}_{\bf k}^{\alpha_1}$ of the $\alpha_1$-pocket. To allow for comparison with our measurements the temperature is set to $T = 0.3$ K such that the system is in the superconducting (SC) state with a calculated gap of $\Delta = 5.5$ meV. The initial parameters are $V = 10$ meV for the impurity potential, $\omega = 8.5$ meV for the boson energy, and $g = 10.9$ meV for the electron-boson coupling strength. As confirmed by our QPI data the calculated  scattering is resonantly enhanced at about 12 meV (dotted line) while it is significantly reduced in an energy range around the Fermi level. b) Momentum cut (lattice constant $a$ as defined in Fig.~\ref{Fig_2}) of the two relevant hole-like bands $\alpha_1$ and $\alpha_2$ in LiFeAs which are used as input parameters of the calculations. The dispersions are taken from the tight-binding model of Ref.~\onlinecite{Saito2015} where the spin orbit coupling in LiFeAs is included. c) Renormalized impurity potential as in (a) but for the temperature $T = 25$ K where the system is in the normal conducting (NC) state. The resonance appears at roughly the same energy as in the superconducting state but is slightly broadened.  Due to higher order effects in combination with the characteristic spin orbit splitting of the $\alpha$-bands the resonance energy deviates approximately by the value $g^2 / \lambda$ ($\approx$ 10 meV) from the boson energy as predicted from the perturbative result \eqref{V_pert}. d) Renormalized impurity potential as in (a,c) but referring to the smaller ${\alpha_2}$-pocket.}

\end{figure}

\subsection{Application to LiFeAs
}
\label{Subsec_theory_appl}

To verify the measured positions of the resonance in LiFeAs in the superconducting as well as in the normal conducting state we have performed a careful analysis of the renormalized scattering potential based on realistic material parameters of LiFeAs. The results are shown in Fig.~\ref{Fig: Theory_2}. For the input of the two relevant hole-like bands $\alpha_1$ and $\alpha_2$ which are predominantly involved in the small-momentum impurity scattering we have used the tight-binding model of Ref.~\onlinecite{Saito2015} which includes also the spin-orbit interaction in LiFeAs. However, to obtain a correct fitting to the small Fermi surfaces of the $\alpha$ bands as measured by ARPES \cite{Borisenko2016} we have used the reduced value $\lambda = 10.5$ meV of the spin orbit coupling parameter instead of the value 50 meV which has been used in Ref.~\onlinecite{Saito2015} to optimally describe  the electron pockets. The corresponding dispersions for  $\lambda = 10.5$ meV are shown in Fig.~\ref{Fig: Theory_2}(b). For the initial boson energy we have used the experimental value $\Omega = 8.5$ meV. The electron-boson coupling is fixed to $g = 10.9$ meV. This particular  value is chosen as a result of a self-consistent superconducting solution of the Hamiltonian $\mathcal{H}_0 +  \mathcal{H}_{eb}$ for the given realistic band structure and boson energy within the approach of Ref.~\onlinecite{Cho2016}. For $g = 10.9$ meV such a self-consistent calculation leads at  temperature $T = 0.3$ K to an $s$-wave gap of $\Delta = 5.5$ meV which is equal to the gap as measured by our STM experiment. 

Taking all the defined parameters as initial conditions we have evaluated separately the PRM renormalization equations for the backscattering impurity potentials $\tilde{V}_{{\bf k},-{\bf k}}^{\alpha_1}$ and $\tilde{V}_{{\bf k},-{\bf k}}^{\alpha_2}$ referring to states ${\bf k}$ and ($-{\bf k}$) within the ${\alpha_1}$- and ${\alpha_2}$-band, respectively.  Panels (a,c) of Fig.~\ref{Fig: Theory_2} show the renormalized impurity scattering potential for the ${\alpha_1}$-band in the superconducting and normal conducting state, respectively. According to the mechanism described above we find a resonantly enhanced scattering potential appearing in form of a maximum in $\tilde{V}_{{\bf k},-{\bf k}}^{\alpha_1}$ at a certain scattering momentum. In the superconducting state, this scattering momentum corresponds to an energy $\tilde{\varepsilon}_{\bf k}^{\alpha_1}$ that is as expected on an approximate level  of $E_{res} \approx \Delta + \Omega \approx 12.8$ meV (dotted line in Fig.~\ref{Fig: Theory_2}(a)) in agreement with the experiment.  In the normal conducting state, however, the resonance is not seen exactly at $\Omega$ as perturbation theory predicts but is shifted to a somewhat larger energy near the position in the superconducting state (dotted line in Fig.~\ref{Fig: Theory_2}(c)). This behavior is also very well consistent with our experimental results (compare Figs.~\ref{Fig_3}, \ref{MDC1}, \ref{MDC2}). The reason for such a shift is the influence of the higher order contributions to the renormalized scattering potential which is discussed in more detail in the following two paragraphs.

The resonance conditions derived above, $\varepsilon_{F}-\varepsilon_{{\bf k}'} = \pm\Omega$ and $\varepsilon_{F}-\varepsilon_{{\bf k}} = \pm\Omega$, are the results of the specific form of the energy denominators in Eq.~\eqref{V_pert} arising as a consequence of perturbation theory.  The condition leads to a singularity in the renormalized impurity potential $\tilde{V}$ which must be removed by the higher order corrections. Schematically, according to Eq.~\eqref{V_pert}, the renormalization equation for the back-scattering impurity potential $\tilde{V}_{{\bf k},-{\bf k}}$ has the following form in second order perturbation theory,
\begin{eqnarray}
\label{Pert_V_schem}
\tilde{V}_{{\bf k},-{\bf k}}^{(2)} = V + \frac{V}{N} \sum_{\bf q} \frac{g^2}{\Delta E_{{\bf k},{\bf q}}\Delta E_{-{\bf k},{\bf q}}} + \dots,
\end{eqnarray}
where $\Delta E_{{\bf k},{\bf q}}$ is an energy denominator which includes the energy difference between electron states and the boson energy. The dots indicate more terms of the same structure. According to the above discussion the conditions $\Delta E_{\pm{\bf k},{\bf q}} = 0$ are responsible for the observed resonance in the impurity scattering for a particular combination of the momentum vectors ${\bf k}$ and ${\bf q}$. 

We now discuss the influence of the higher order contributions to the perturbative result \eqref{Pert_V_schem}.
According to the method of continued fractions (see for example Ref.~\onlinecite{Horacek1983}) which can be considered for the solution of integral equations the higher order correction to an arbitrary  energy denominator $\Delta E$ is $\Delta E + g^2/\Delta E'$ where the correction $g^2 / \Delta E'$ must be continued to all denominators up to infinite order, i.~e.~$\Delta E' = \Delta E +g^2/\Delta E''$, $\Delta E'' = \Delta E + g^2 / \dots$. Thus, the corresponding higher-order correction of Eq.~\eqref{Pert_V_schem} reads,
\begin{eqnarray}
\label{Corr_V_schem} 
&& \tilde{V}_{{\bf k},-{\bf k}} = V  + \\ 
&&  \frac{V}{N} \sum_{\bf q} \frac{g^2}{\left( \Delta E_{{\bf k},{\bf q}} +  \frac{1}{N} \sum_{{\bf q}'}\frac{g^2}{\Delta E_{{\bf k},{\bf q}'} + \dots} \right) \Big(\dots \Big)} + \dots, \nonumber
\end{eqnarray}
where the factor $(\dots)$ denotes an equivalent factor with ${\bf k}$ replaced by  $-{\bf k}$.
As can be seen easily from the change of the energy denominator in Eq.~\eqref{Corr_V_schem} such a correction leads immediately to a shift of the resonance condition from $\Delta E_{\pm{\bf k},{\bf q}} = 0$ (as in perturbation theory) to 
\begin{eqnarray}
\label{Corr_Res}
\Delta E_{\pm{\bf k},{\bf q}} +  \frac{1}{N} \sum_{{\bf q}'}\frac{g^2}{\Delta E_{\pm{\bf k},{\bf q}'} + \dots} = 0,
\end{eqnarray}
where the energy denominator has the particular form $\Delta E_{\pm{\bf k},{\bf q}} = \varepsilon_{\pm {\bf k}} -  \varepsilon_{\pm {\bf k} + {\bf q}} + \Omega$ in a one-band system. Eq.~\eqref{Corr_Res} enables to estimate the higher-order correction to the perturbation theory value of the resonance energy. We start with the simplest case of a usual metal.  In this case $\Delta E_{\pm{\bf k},{\bf q}'}$ is of the order of (eV) for most of the momentum vectors ${\bf q}'$ in the momentum summation since for a given $\pm{\bf k}$ the energy difference $\varepsilon_{\pm {\bf k}} -  \varepsilon_{\pm {\bf k}+ {\bf q}'}$ combines high-energy states for a macroscopic amount (order of $N$) of ${\bf q}'$ points. Thus, according to Eq.~\eqref{Corr_Res}, for a usual one-band metal the correction is of the order $g^2/\mbox{(eV)}$ which is usually a very small value. For such a material the perturbation theory is valid. However, the situation changes in the superconducting state where due to the presence of the superconducting gap $\Delta$ a macroscopic number of states gives rise to energy transitions $\Delta E_{\pm{\bf k},{\bf q}'}$ of the order of $\Delta$. In  case of superconducting LiFeAs, the correction to the resonance condition in Eq.~\eqref{Corr_Res} is also of the order of $\Delta$ since $g \approx 10$ meV. This explains the shift of the resonance in the superconducting state from $\Omega$ to roughly $\Omega + g^2/\Delta \approx \Omega + \Delta$. 

In the normal conducting state of LiFeAs we would expect a shift according to the same arguments. This is due to the specific band structure in LiFeAs. As can be seen in Fig.~\ref{Fig: Theory_2}(b) there is a large density of states around the $\Gamma$ point where the $\alpha$ bands have maximum values. However, in this region the spin orbit splitting is significant and also of the  order of 10 meV (as the superconducting gap). Thus, the correction  to the perturbative resonance condition in the normal conducting state is  of the same order as in the superconducting state, which is clearly seen in a comparison between Figs.~\ref{Fig: Theory_2}(a) and \ref{Fig: Theory_2}(c).

Furthermore, one can recognize  a characteristic depletion of $\tilde{V}_{{\bf k},-{\bf k}}^{\alpha_1}$ in a region around the Fermi level. This feature is again  a consequence of the specific band structure and can be understood as follows. Due to the relatively strong $k_z$ dispersion of both bands (compare solid and dashed lines in Fig.~\ref{Fig: Theory_2}(b)) the top of the ${\alpha_1}$-band ranges from about 8 meV to 16 meV. Thus, for scattering energies below about 8 meV the sign of the dominant energy denominators, which involve states around the top of the band, can change. 

Moreover, at negative energies no further resonance is found for the $\alpha_1$-band which is also in agreement with the experiment. Instead, such a resonance is found in a much weaker form on the $\alpha_2$-band as shown in Fig.~\ref{Fig: Theory_2}(d). This is the reason why in our QPI measurements the feature at negative bias voltage appears at similar momentum as the main resonance feature on the positive side.  This observation suggests that the involved boson, which leads to the exaggerated impurity  scattering and at the same time mediates the superconducting pairing, must have a very small momentum ${\bf q}$.

In our QPI measurements the intensity of the resonance at positive energy is exaggerated in comparison with the usual QPI at larger momentum or negative energy. As discussed in Fig.~\ref{Fig: Theory_2} such an amplification is clearly seen also in our theoretical model by a significant variation of $\tilde{V}_{{\bf k},-{\bf k}}^{\alpha_1}$  by a factor of around 2. This amount of variation is sufficient to explain the observed resonance behavior in the QPI measurements for the following reasons. Firstly, note that the QPI intensity is not only determined by the renormalized scattering potential but also by the density of states which is enhanced near the band maximum of the $\alpha$-bands. Moreover, additional contributions to the QPI are given by multiple scattering processes, which become particularly important in the presence of a strong scattering potential. Thus, a consistent calculation of the QPI intensity requires in this case the self-consistent inclusion of higher order scattering processes by a standard $t$-matrix approach. Such  treatments are known to  boost  the QPI intensity particularly at energies around a resonance. These reasons altogether lead to the conclusion that our calculated variation of the scattering potential can fully explain our measured resonance behavior in the QPI of LiFeAs.

\section{Experimental details}

\label{Sec_Methods}

\begin{figure}[ht]

\centering

\includegraphics[width=\columnwidth]{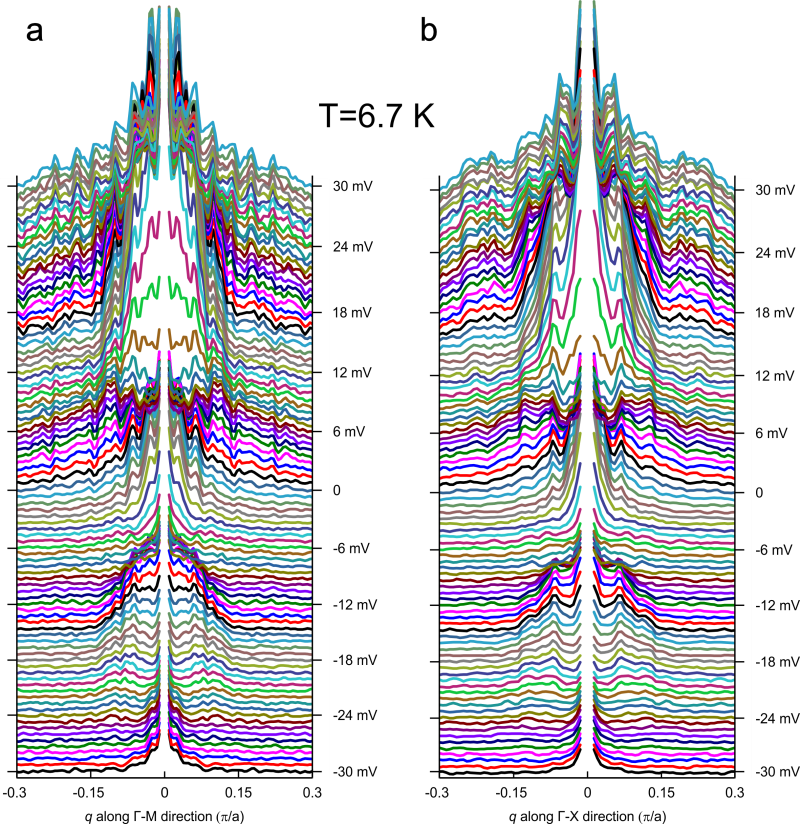}
 \caption{Waterfall representation of the FT-STS data taken at 6.7~K. (\textbf{a}) along  $\Gamma-M$, (\textbf{b}) along $\Gamma-X$.}
\label{waterfall} 
\end{figure}

\begin{figure}[ht]

\centering

\includegraphics[width=\columnwidth]{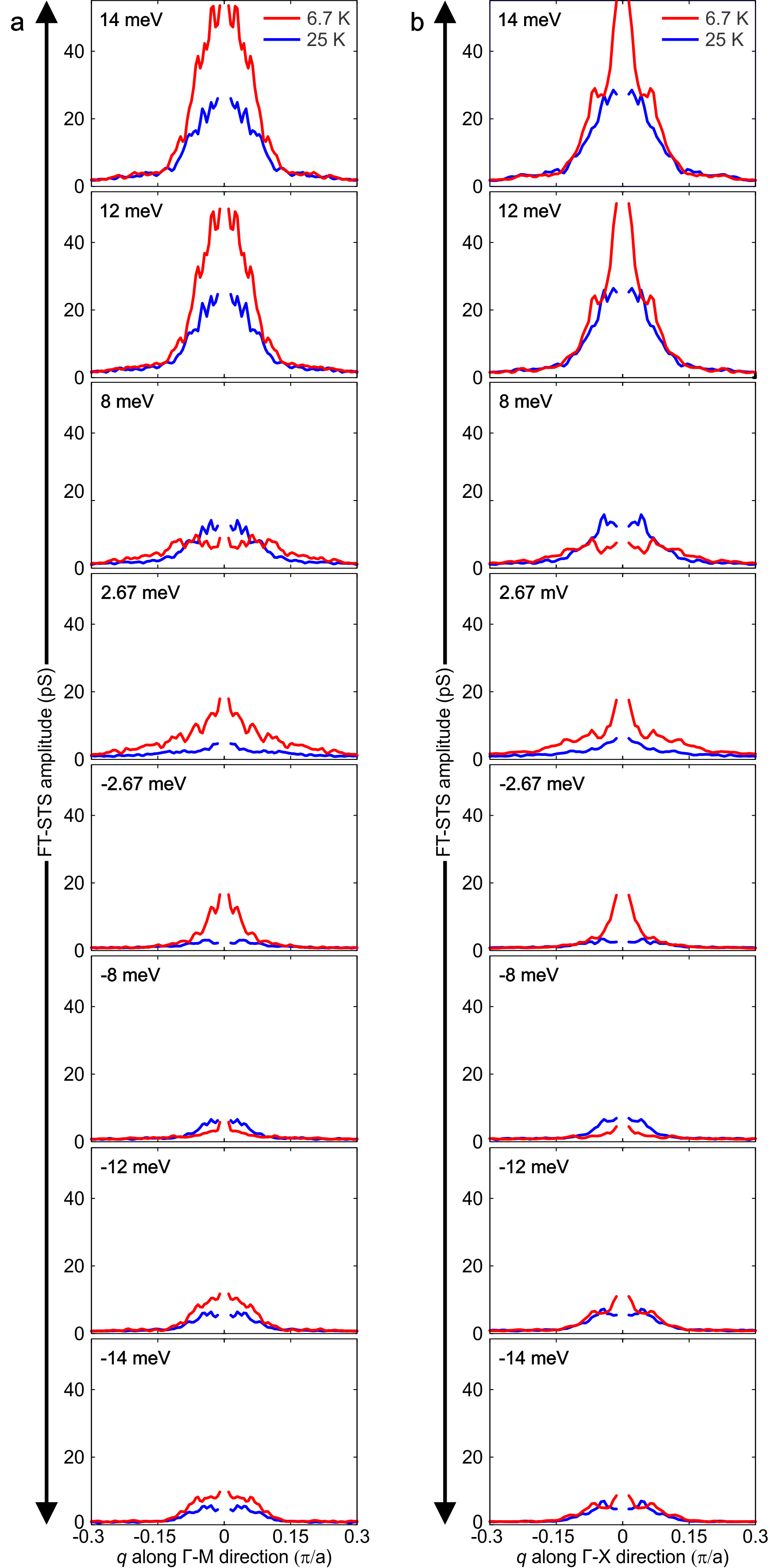}
 \caption{Amplitude of the FT-STS data for selected energies at 6.7~K and at 25~K. Columns (a) and (b) show the data along the $\Gamma-M$ and $\Gamma-X$ directions, respectively.}
\label{linecuts}
\end{figure}
\subsection{Sample preparation}

Single crystals of stoichiometric LiFeAs have been grown using the self-flux method as described in Ref.~\onlinecite{Morozov2010}. In order to ensure stoichiometry and homogeneity of the sample, we confirmed the $^{75}$As NQR frequency and line width of the sample as 21.561~$\pm$~0.001~MHz and 31~$\pm$~1~kHz, respectively \cite{Nag2016}. Since LiFeAs is highly air sensitive, these steps, and the mounting of the sample into our STM have been performed in Ar atmosphere.

\subsection{Scanning tunneling microscopy/spectroscopy measurements}

The STM measurements are carried out in two home-built low-temperature scanning tunneling microscopes using a tungsten tip. One of the microscopes is optimized for QPI data acquisition at variable temperatures \cite{Schlegel2014}. All QPI  and point spectroscopy data at temperatures between 6.7~K and 25~K have been obtained with this instrument on one single crystal of LiFeAs. The other \cite{Salazar2018} has been used for measuring the 300~mK point spectroscopy data on another LiFeAs crystal.

Atomically flat LiFeAs surfaces were obtained by cleaving the crystal inside the STM in cryogenic vacuum or ultra-high vacuum. For all tunneling conductance spectra, we used a lock-in amplifier with a modulation of 0.4~mV$_\mathrm{rms}$ at 1.1111 kHz. 
Conductance maps are taken with a grid size of $256\times256$ pixels.
All the spectroscopic maps are taken with stabilization condition of $U_\mathrm{bias}$=-50~mV and $I_T$=600~pA. Each spectroscopic map is measured over the energy range between $\pm$30~mV with consecutive energy point spacing of 0.67~mV. The total time for acquiring one spectroscopic map was about 3.5 days.
Prior to each spectroscopic map measurements, the microscope was stabilized at the respective temperature for a sufficient time until a stable drift of the tip with respect to the sample was reached. At base temperature (6.7~K) the drift was immeasurably small, whereas at 25~K the drift was lower than 2$a$ per day.

\begin{figure*}[t]
\centering
\includegraphics[width=0.9\textwidth]{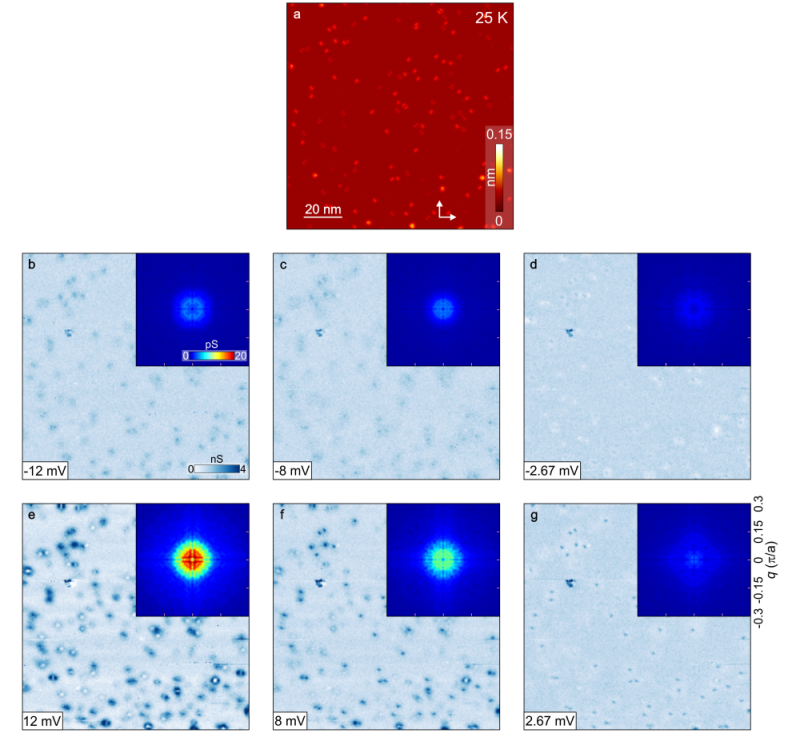}
\caption{Experimental data in the normal conducting state at $T=25$~K. (a) Representative surface topography ($U_\mathrm{bias}= -50$~mV, $I = 100$~pA) of the sample surface. The directions of the shortest Fe-Fe distance are indicated by arrows. (b-g) Real space conductance map at selected energies $eU_\mathrm{bias}=\pm12$~meV, $\pm8$~meV, and $\pm2.67$~meV. The conductance map is taken at the same area shown in (a).  The Fourier transformation of the real space conductance map data is shown in the corresponding insets.}
\label{Fig: bz25}
\end{figure*}

\subsection{Data processing}
\label{FT_method}
The FT-STS data is calculated as the amplitude of the fast Fourier transform of the AC part of each energy slice of the real-space spectroscopy map. Symmetrized FT-STS data sets and images are subsequently achieved by symmetrizing the raw FT-STS data along both the two lattice high symmetry directions ($\Gamma-X$ and $\Gamma-M$) (as shown in Fig.~\ref{Fig_2}(c)).
In order to enhance the contrast of the QPI pattern, we have applied a ($5\times5$) linear convolvement to the symmetrized QPI data, from which the line cuts in Figs.~\ref{Fig_3} (a-d) have been derived. 

\section{Additional experimental data} 
\label{Additional_exp}

\subsection{Superconducting state at 6.7~K}

The whole data set at base temperature (6.7~K), of which Fig.~\ref{Fig_2} shows selected energy slices, is visualized in the Movie S1 \cite{SM}. 
% It shows both the real space and Fourier transformed data.
In order to further visualize the energy and momentum dependence of the amplitude of the FT-STS data we show, complementary to Figs.~\ref{Fig_3}(a) and \ref{Fig_3}(b), in Fig.~\ref{waterfall} a waterfall representation of these data along the $\Gamma-M$ direction (panel (a)) and along the $\Gamma-X$ direction (panel (b)).

Fig.~\ref{linecuts} depicts the amplitude of the FT-STS data in the superconducting state at 6.7~K for selected energies in line-cuts along  $\Gamma-M$ and $\Gamma-X$ in order to highlight the strong enhancement of the amplitude at energies larger than about 10~meV, and to demonstrate the decay of the amplitude as a function of $q$.

\subsection{Normal conducting state at 25~K}

\begin{figure}
\centering
\includegraphics[width=\columnwidth]{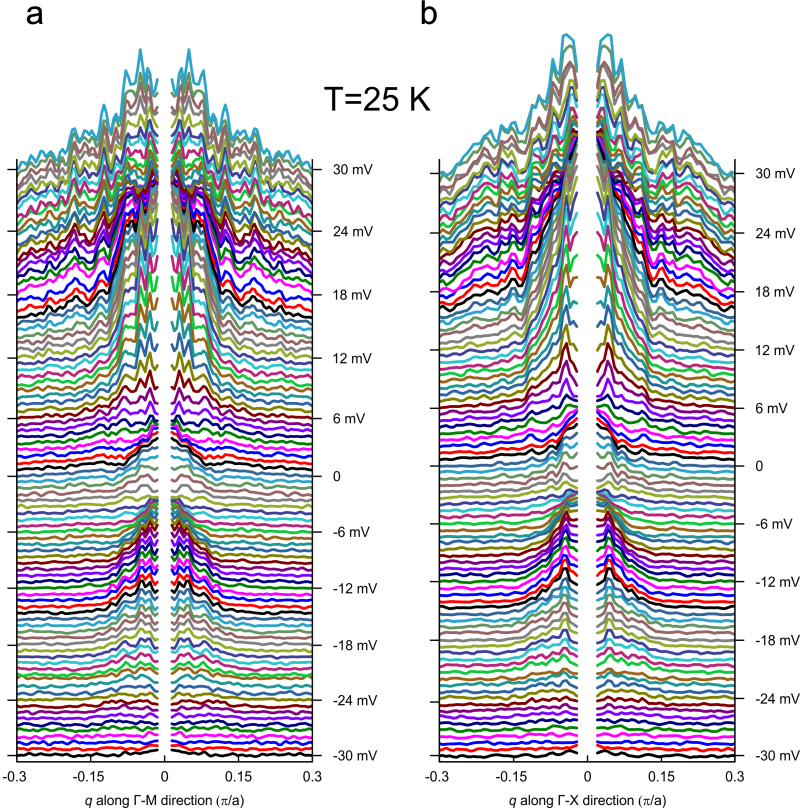}
 \caption{Waterfall representation of the FT-STS data taken at 25~K. (a) along  $\Gamma-M$, (b) along $\Gamma-X$.}
 \label{waterfall_ns}
\end{figure}

Fig.~\ref{Fig: bz25} shows the topography and selected energy slices in real space and the corresponding Fourier transformed data of the conductance map measured in the normal state at 25~K. Based on these results, the low temperature (6.7~K) data in Fig.~\ref{linecuts} are complemented by analogous data for the normal conducting state at 25~K. Evidently, the FT-STS amplitude at energies larger than 10~meV, whereas the amplitude enhancement due the impurity bound state has vanished ($\pm2.67$~meV).
In order to further visualize the energy and momentum dependence of the amplitude of the FT-STS data we show, complementary to Figs.~\ref{Fig_3}(c) and \ref{Fig_3}(d), in Fig.~\ref{waterfall_ns} a waterfall representation of the FT-STS data along the $\Gamma-M$ direction (panel (a)) and along the $\Gamma-X$ direction (panel (b)).

Fig.~\ref{linecuts} depicts the amplitude of the FT-STS data for selected energies in order to highlight the strong enhancement of the amplitude at energies larger than about 10~meV, and to demonstrate the decay of the amplitude as a function of $q$. An overall smaller amplitude as compared to that at 6.7~K is evident.

\subsection{Temperature dependent FT-STS data}
\label{Temp_data}
\begin{figure}[t]
\centering
\includegraphics[width=0.95\columnwidth]{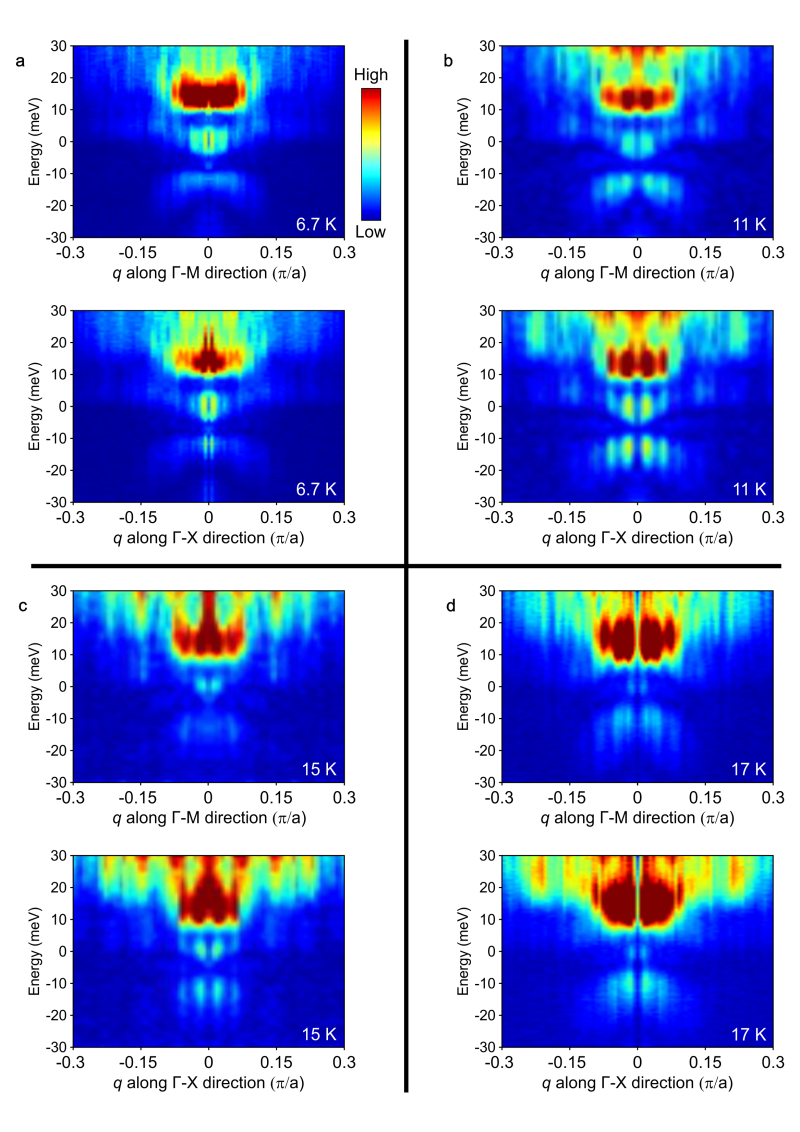}
 \caption{Energy-momentum dependence of FT-STS data along high-symmetry directions. Panels (a-d) correspond to 6.7~K, 11~K, 15~K, and 17~K, respectively.}
\label{MDC1}
\end{figure}

\begin{figure}[p]
\centering
\includegraphics[width=0.95\columnwidth]{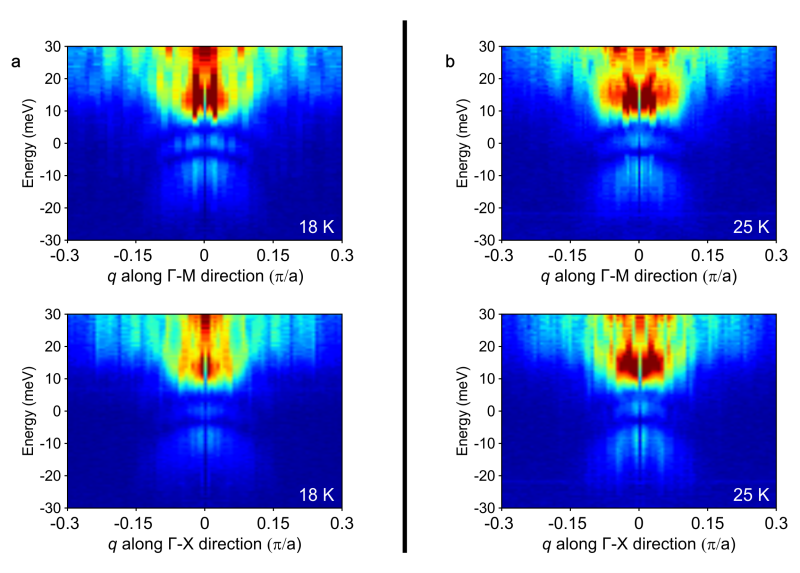}
 \caption{Energy-momentum dependence of FT-STS data along high-symmetry directions. Panels (a) and (b) correspond to  18~K and 25~K, respectively.}
\label{MDC2}
\end{figure}

\begin{figure}[p]
\centering
\includegraphics[width=\columnwidth]{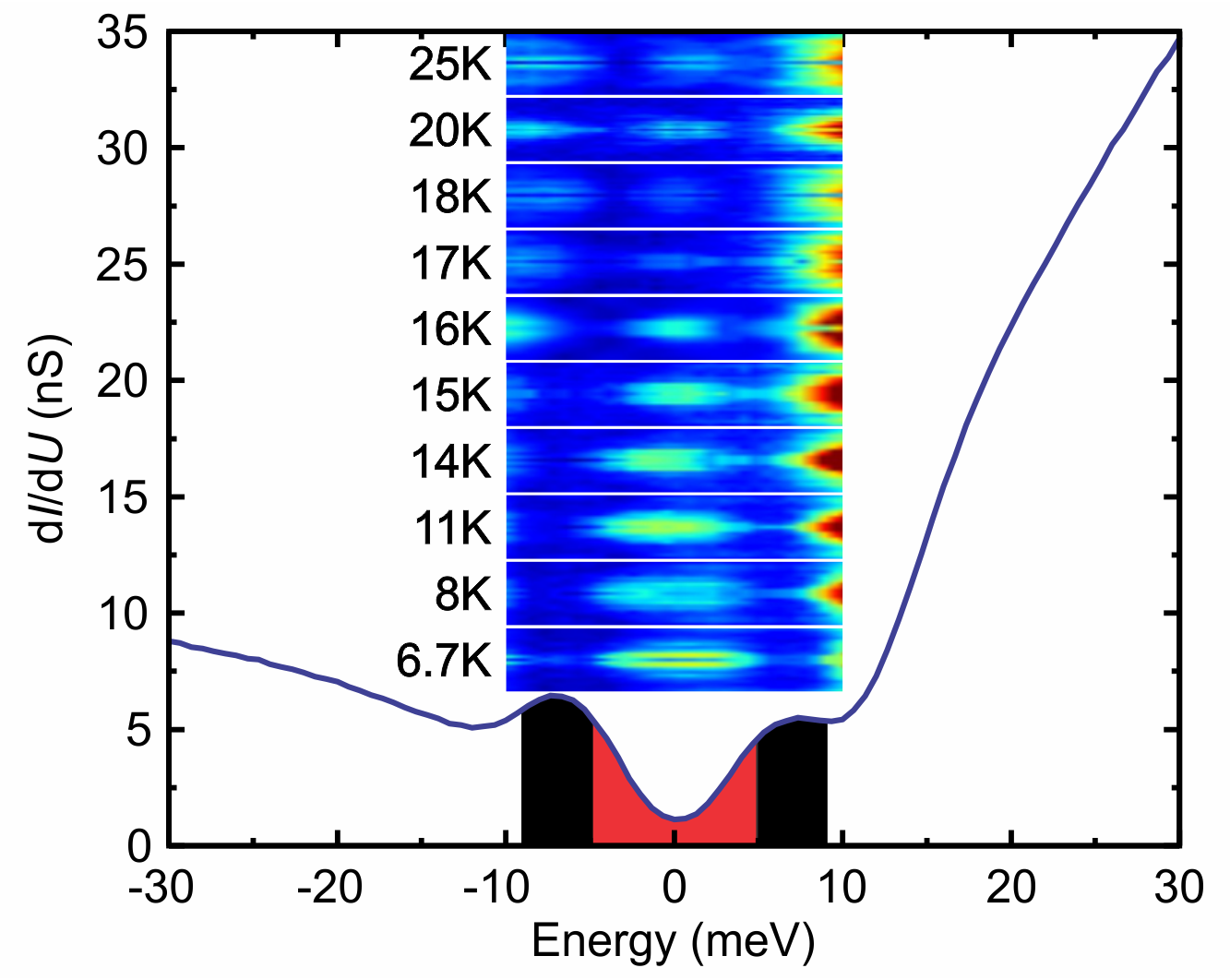}
 \caption{Temperature evolution of feature i). Representative d$I$/d$U$ spectrum on the bare surface at 6.7~K highlighting the size of $\Delta_1$ (red bar) in comparison to feature i). Inset: Zoom-in into the area spanned by momentum $q=\pm0.086\pi/a$  and energy $eU_\mathrm{bias}=\pm$10~meV for each temperature of the line-cuts shown in Figs.~\ref{MDC1} and \ref{MDC2} and for further temperatures along the $\Gamma-M$ direction.}
 \label{feature1}
\end{figure}

$\mathrm{d}I/\mathrm{d}U$ maps of  110~nm~$\times$~110~nm areas have been measured in the temperature range between 6.7~K and 25~K at 10 specific temperatures. The resulting energy-momentum dependence of the FT-STS data at selected energies is shown in Fig.~\ref{MDC1} and Fig.~\ref{MDC2} as line-cuts along high-symmetry directions analogous to Figs.~\ref{Fig_3}(a) and \ref{Fig_3}(b). These data show quite clearly that feature ii) remains present at all temperatures and even in the normal conducting state, whereas feature i) vanishes at the critical temperature $T_c\approx18$~K. The latter is further visualized in Fig.~\ref{feature1}.

\begin{figure}[p]
\centering
\includegraphics[width=\columnwidth]{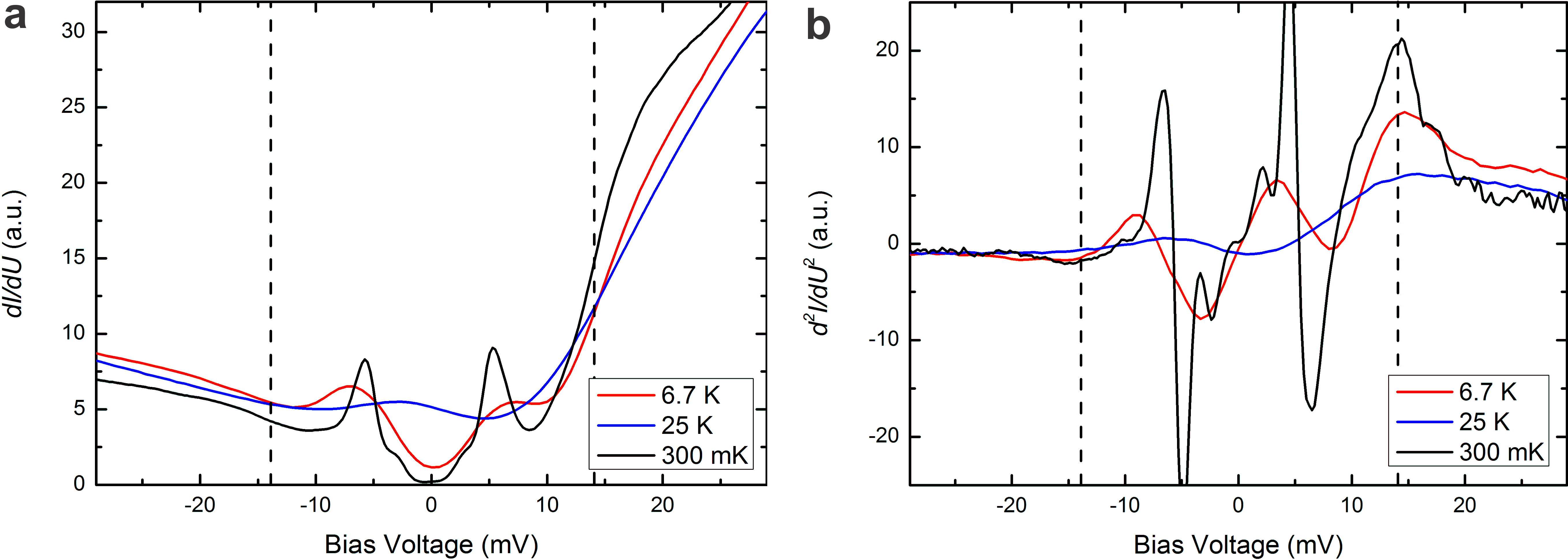}
 \caption{Low-temperature point spectroscopy. Comparison of 300~mK point spectroscopy data with those shown in Figs.~\ref{Fig_4}(d,e). (a) $\mathrm{d}I/\mathrm{d}U$ data at 300~mK, 6.7~K, and 25~K. (b) Second derivative  $\mathrm{d}^2I/\mathrm{d}U^2$ for the same temperatures.}
 \label{low-t-spec}
\end{figure}

\begin{figure}[h]
\centering
\includegraphics[width=\columnwidth]{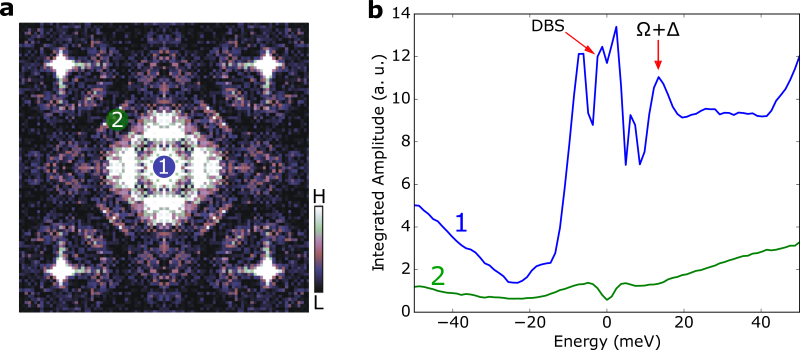}
 \caption{Comparison of feature ii) with QPI signals at larger $q$. (a) Representative FT-STS data (the same as published in Ref.~\onlinecite{Haenke2012}) at $-14.1$~meV. The small-$q$ area which is at the focus of this work is indicated by the shaded area '1'. The shaded area '2' is located around those $q$ where intraband scattering of the $\gamma$-band is located (labeled ${\bf q}_4$ in Ref.~\onlinecite{Haenke2012}). Apparently, this intraband scattering generates a particularly sharp and intense QPI signal, which is caused by a significant nesting of the band (cf. Fig.~\ref{FS_schem}). We integrate the amplitudes at areas '1' and '2' and plot the integrated amplitude in panel (b) as a function of energy. Clearly, the integrated amplitude around '2' remains significantly lower than that of area '1'. In particular, a pronounced peak at $\Omega+\Delta$  corresponding to feature ii) of the current study is visible, in addition to clear signatures due to defect bound states (DBS).
 }
 \label{Fig_supp_old_QPI}
\end{figure}

\subsection{Low-temperature point spectroscopy}
Fig.~\ref{low-t-spec} shows a comparison of the 300~mK point spectroscopy data of Fig.~\ref{Fig_2}(a) with those shown in Figs. \ref{Fig_4}(d,e). The $\mathrm{d}I/\mathrm{d}U$ data at 300~mK apparently are a systematic low-temperature evolution of the data at 6.7~K, since the coherence peaks of the large gap $\Delta_1$ and of the small gap $\Delta_2$ become clearly discernible (panel (a) of Fig.~\ref{low-t-spec}). The second derivative  $\mathrm{d}^2I/\mathrm{d}U^2$ shown in Fig.~\ref{low-t-spec}(b) underpins the sharpening of the spectral features at $|eU|\lesssim\Delta_1$. In contrast, the width of the step-like increase in  $\mathrm{d}I/\mathrm{d}U$ at about 14~mV reduces only by a small amount upon cooling from 6.7~K to 300~mK, as is revealed by the corresponding peaks in $\mathrm{d}^2I/\mathrm{d}U^2$.

\section{Comparison of feature ii) with standard QPI at larger $q$}
\label{Standard_QPI}
Fig.~\ref{Fig_supp_old_QPI} shows a comparison of the integrated amplitude of feature ii) with that of standard QPI. For this comparison, we analyzed our previous LiFeAs data \cite{Haenke2012} which include such large-$q$ QPI information. The inspection of these data clearly confirm the presence of feature ii) and furthermore reveal that feature ii) is significantly more intense than standard QPI.

\end{appendix}
\newpage

% \bibliography{ROFFeAs,stm,QPI,notes}

\begin{thebibliography}{56}
\expandafter\ifx\csname natexlab\endcsname\relax\def\natexlab#1{#1}\fi
\expandafter\ifx\csname bibnamefont\endcsname\relax
  \def\bibnamefont#1{#1}\fi
\expandafter\ifx\csname bibfnamefont\endcsname\relax
  \def\bibfnamefont#1{#1}\fi
\expandafter\ifx\csname citenamefont\endcsname\relax
  \def\citenamefont#1{#1}\fi
\expandafter\ifx\csname url\endcsname\relax
  \def\url#1{\texttt{#1}}\fi
\expandafter\ifx\csname urlprefix\endcsname\relax\def\urlprefix{URL }\fi
\providecommand{\bibinfo}[2]{#2}
\providecommand{\eprint}[2][]{\url{#2}}

\bibitem[{\citenamefont{McMillan and Rowell}(1965)}]{McMillan1965}
\bibinfo{author}{\bibfnamefont{W.~L.} \bibnamefont{McMillan}} \bibnamefont{and}
  \bibinfo{author}{\bibfnamefont{J.~M.} \bibnamefont{Rowell}},
  \bibinfo{journal}{Phys. Rev. Lett.} \textbf{\bibinfo{volume}{14}},
  \bibinfo{pages}{108} (\bibinfo{year}{1965}),
  \urlprefix\url{https://link.aps.org/doi/10.1103/PhysRevLett.14.108}.

\bibitem[{\citenamefont{Scalapino et~al.}(1966)\citenamefont{Scalapino,
  Schrieffer, and Wilkins}}]{Scalapino1966}
\bibinfo{author}{\bibfnamefont{D.~J.} \bibnamefont{Scalapino}},
  \bibinfo{author}{\bibfnamefont{J.~R.} \bibnamefont{Schrieffer}},
  \bibnamefont{and} \bibinfo{author}{\bibfnamefont{J.~W.}
  \bibnamefont{Wilkins}}, \bibinfo{journal}{Phys. Rev.}
  \textbf{\bibinfo{volume}{148}}, \bibinfo{pages}{263} (\bibinfo{year}{1966}),
  \urlprefix\url{http://link.aps.org/doi/10.1103/PhysRev.148.263}.

\bibitem[{\citenamefont{Hudson et~al.}(1999)\citenamefont{Hudson, Pan, Gupta,
  Ng, and Davis}}]{Hudson1999}
\bibinfo{author}{\bibfnamefont{E.~W.} \bibnamefont{Hudson}},
  \bibinfo{author}{\bibfnamefont{S.~H.} \bibnamefont{Pan}},
  \bibinfo{author}{\bibfnamefont{A.~K.} \bibnamefont{Gupta}},
  \bibinfo{author}{\bibfnamefont{K.-W.} \bibnamefont{Ng}}, \bibnamefont{and}
  \bibinfo{author}{\bibfnamefont{J.~C.} \bibnamefont{Davis}},
  \bibinfo{journal}{Science} \textbf{\bibinfo{volume}{285}},
  \bibinfo{pages}{88} (\bibinfo{year}{1999}), ISSN \bibinfo{issn}{0036-8075},
  \urlprefix\url{http://science.sciencemag.org/content/285/5424/88}.

\bibitem[{\citenamefont{Jenkins et~al.}(2009)\citenamefont{Jenkins, Fasano,
  Berthod, Maggio-Aprile, Piriou, Giannini, Hoogenboom, Hess, Cren, and
  Fischer}}]{Jenkins2009}
\bibinfo{author}{\bibfnamefont{N.}~\bibnamefont{Jenkins}},
  \bibinfo{author}{\bibfnamefont{Y.}~\bibnamefont{Fasano}},
  \bibinfo{author}{\bibfnamefont{C.}~\bibnamefont{Berthod}},
  \bibinfo{author}{\bibfnamefont{I.}~\bibnamefont{Maggio-Aprile}},
  \bibinfo{author}{\bibfnamefont{A.}~\bibnamefont{Piriou}},
  \bibinfo{author}{\bibfnamefont{E.}~\bibnamefont{Giannini}},
  \bibinfo{author}{\bibfnamefont{B.~W.} \bibnamefont{Hoogenboom}},
  \bibinfo{author}{\bibfnamefont{C.}~\bibnamefont{Hess}},
  \bibinfo{author}{\bibfnamefont{T.}~\bibnamefont{Cren}}, \bibnamefont{and}
  \bibinfo{author}{\bibfnamefont{O.}~\bibnamefont{Fischer}},
  \bibinfo{journal}{Phys. Rev. Lett.} \textbf{\bibinfo{volume}{103}},
  \bibinfo{pages}{227001} (\bibinfo{year}{2009}),
  \urlprefix\url{https://link.aps.org/doi/10.1103/PhysRevLett.103.227001}.

\bibitem[{\citenamefont{Wang et~al.}(2013{\natexlab{a}})\citenamefont{Wang,
  Yang, Fang, Shen, Wang, Shan, Zhang, Dai, and Wen}}]{Wang2013d}
\bibinfo{author}{\bibfnamefont{Z.}~\bibnamefont{Wang}},
  \bibinfo{author}{\bibfnamefont{H.}~\bibnamefont{Yang}},
  \bibinfo{author}{\bibfnamefont{D.}~\bibnamefont{Fang}},
  \bibinfo{author}{\bibfnamefont{B.}~\bibnamefont{Shen}},
  \bibinfo{author}{\bibfnamefont{Q.-H.} \bibnamefont{Wang}},
  \bibinfo{author}{\bibfnamefont{L.}~\bibnamefont{Shan}},
  \bibinfo{author}{\bibfnamefont{C.}~\bibnamefont{Zhang}},
  \bibinfo{author}{\bibfnamefont{P.}~\bibnamefont{Dai}}, \bibnamefont{and}
  \bibinfo{author}{\bibfnamefont{H.-H.} \bibnamefont{Wen}},
  \bibinfo{journal}{Nat Phys} \textbf{\bibinfo{volume}{9}}, \bibinfo{pages}{42}
  (\bibinfo{year}{2013}{\natexlab{a}}), ISSN \bibinfo{issn}{1745-2473},
  \urlprefix\url{http://dx.doi.org/10.1038/nphys2478}.

\bibitem[{\citenamefont{Song et~al.}(2014)\citenamefont{Song, Wang, Jiang, Li,
  Wang, He, Chen, Hoffman, Ma, and Xue}}]{Song2014}
\bibinfo{author}{\bibfnamefont{C.-L.} \bibnamefont{Song}},
  \bibinfo{author}{\bibfnamefont{Y.-L.} \bibnamefont{Wang}},
  \bibinfo{author}{\bibfnamefont{Y.-P.} \bibnamefont{Jiang}},
  \bibinfo{author}{\bibfnamefont{Z.}~\bibnamefont{Li}},
  \bibinfo{author}{\bibfnamefont{L.}~\bibnamefont{Wang}},
  \bibinfo{author}{\bibfnamefont{K.}~\bibnamefont{He}},
  \bibinfo{author}{\bibfnamefont{X.}~\bibnamefont{Chen}},
  \bibinfo{author}{\bibfnamefont{J.~E.} \bibnamefont{Hoffman}},
  \bibinfo{author}{\bibfnamefont{X.-C.} \bibnamefont{Ma}}, \bibnamefont{and}
  \bibinfo{author}{\bibfnamefont{Q.-K.} \bibnamefont{Xue}},
  \bibinfo{journal}{Phys. Rev. Lett.} \textbf{\bibinfo{volume}{112}},
  \bibinfo{pages}{057002} (\bibinfo{year}{2014}),
  \urlprefix\url{http://link.aps.org/doi/10.1103/PhysRevLett.112.057002}.

\bibitem[{\citenamefont{Chi et~al.}(2012)\citenamefont{Chi, Grothe, Liang,
  Dosanjh, Hardy, Burke, Bonn, and Pennec}}]{Chi2012}
\bibinfo{author}{\bibfnamefont{S.}~\bibnamefont{Chi}},
  \bibinfo{author}{\bibfnamefont{S.}~\bibnamefont{Grothe}},
  \bibinfo{author}{\bibfnamefont{R.}~\bibnamefont{Liang}},
  \bibinfo{author}{\bibfnamefont{P.}~\bibnamefont{Dosanjh}},
  \bibinfo{author}{\bibfnamefont{W.~N.} \bibnamefont{Hardy}},
  \bibinfo{author}{\bibfnamefont{S.~A.} \bibnamefont{Burke}},
  \bibinfo{author}{\bibfnamefont{D.~A.} \bibnamefont{Bonn}}, \bibnamefont{and}
  \bibinfo{author}{\bibfnamefont{Y.}~\bibnamefont{Pennec}},
  \bibinfo{journal}{Phys. Rev. Lett.} \textbf{\bibinfo{volume}{109}},
  \bibinfo{pages}{087002} (\bibinfo{year}{2012}),
  \urlprefix\url{http://link.aps.org/doi/10.1103/PhysRevLett.109.087002}.

\bibitem[{\citenamefont{Nag et~al.}(2016)\citenamefont{Nag, Schlegel, Baumann,
  Grafe, Beck, Wurmehl, B\"uchner, and Hess}}]{Nag2016}
\bibinfo{author}{\bibfnamefont{P.~K.} \bibnamefont{Nag}},
  \bibinfo{author}{\bibfnamefont{R.}~\bibnamefont{Schlegel}},
  \bibinfo{author}{\bibfnamefont{D.}~\bibnamefont{Baumann}},
  \bibinfo{author}{\bibfnamefont{H.-J.} \bibnamefont{Grafe}},
  \bibinfo{author}{\bibfnamefont{R.}~\bibnamefont{Beck}},
  \bibinfo{author}{\bibfnamefont{S.}~\bibnamefont{Wurmehl}},
  \bibinfo{author}{\bibfnamefont{B.}~\bibnamefont{B\"uchner}},
  \bibnamefont{and} \bibinfo{author}{\bibfnamefont{C.}~\bibnamefont{Hess}},
  \bibinfo{journal}{Scientific Reports} \textbf{\bibinfo{volume}{6}},
  \bibinfo{pages}{27926} (\bibinfo{year}{2016}),
  \urlprefix\url{http://dx.doi.org/10.1038/srep27926}.

\bibitem[{\citenamefont{Hlobil et~al.}(2017)\citenamefont{Hlobil, Jandke,
  Wulfhekel, and Schmalian}}]{Hlobil2017}
\bibinfo{author}{\bibfnamefont{P.}~\bibnamefont{Hlobil}},
  \bibinfo{author}{\bibfnamefont{J.}~\bibnamefont{Jandke}},
  \bibinfo{author}{\bibfnamefont{W.}~\bibnamefont{Wulfhekel}},
  \bibnamefont{and}
  \bibinfo{author}{\bibfnamefont{J.}~\bibnamefont{Schmalian}},
  \bibinfo{journal}{Phys. Rev. Lett.} \textbf{\bibinfo{volume}{118}},
  \bibinfo{pages}{167001} (\bibinfo{year}{2017}),
  \urlprefix\url{https://link.aps.org/doi/10.1103/PhysRevLett.118.167001}.

\bibitem[{\citenamefont{Johnston}(2010)}]{Johnston2010}
\bibinfo{author}{\bibfnamefont{D.~C.} \bibnamefont{Johnston}},
  \bibinfo{journal}{Advances in Physics} \textbf{\bibinfo{volume}{59}},
  \bibinfo{pages}{803} (\bibinfo{year}{2010}),
  \urlprefix\url{http://www.informaworld.com/10.1080/00018732.2010.513480}.

\bibitem[{\citenamefont{Fernandes et~al.}(2014)\citenamefont{Fernandes,
  Chubukov, and Schmalian}}]{Fernandes2014}
\bibinfo{author}{\bibfnamefont{R.~M.} \bibnamefont{Fernandes}},
  \bibinfo{author}{\bibfnamefont{A.~V.} \bibnamefont{Chubukov}},
  \bibnamefont{and}
  \bibinfo{author}{\bibfnamefont{J.}~\bibnamefont{Schmalian}},
  \bibinfo{journal}{Nat Phys} \textbf{\bibinfo{volume}{10}},
  \bibinfo{pages}{97} (\bibinfo{year}{2014}), ISSN \bibinfo{issn}{1745-2473},
  \urlprefix\url{http://dx.doi.org/10.1038/nphys2877}.

\bibitem[{\citenamefont{Mazin et~al.}(2008)\citenamefont{Mazin, Singh,
  Johannes, and Du}}]{Mazin2008}
\bibinfo{author}{\bibfnamefont{I.~I.} \bibnamefont{Mazin}},
  \bibinfo{author}{\bibfnamefont{D.~J.} \bibnamefont{Singh}},
  \bibinfo{author}{\bibfnamefont{M.~D.} \bibnamefont{Johannes}},
  \bibnamefont{and} \bibinfo{author}{\bibfnamefont{M.~H.} \bibnamefont{Du}},
  \bibinfo{journal}{Phys. Rev. Lett.} \textbf{\bibinfo{volume}{101}},
  \bibinfo{pages}{057003} (\bibinfo{year}{2008}),
  \urlprefix\url{http://dx.doi.org/10.1103/PhysRevLett.101.057003}.

\bibitem[{\citenamefont{Kontani and Onari}(2010)}]{Kontani2010}
\bibinfo{author}{\bibfnamefont{H.}~\bibnamefont{Kontani}} \bibnamefont{and}
  \bibinfo{author}{\bibfnamefont{S.}~\bibnamefont{Onari}},
  \bibinfo{journal}{Phys. Rev. Lett.} \textbf{\bibinfo{volume}{104}},
  \bibinfo{pages}{157001} (\bibinfo{year}{2010}),
  \urlprefix\url{http://link.aps.org/doi/10.1103/PhysRevLett.104.157001}.

\bibitem[{\citenamefont{Inosov et~al.}(2010)\citenamefont{Inosov, Park,
  Bourges, Sun, Sidis, Schneidewind, Hradil, Haug, Lin, Keimer
  et~al.}}]{Inosov2010a}
\bibinfo{author}{\bibfnamefont{D.~S.} \bibnamefont{Inosov}},
  \bibinfo{author}{\bibfnamefont{J.~T.} \bibnamefont{Park}},
  \bibinfo{author}{\bibfnamefont{P.}~\bibnamefont{Bourges}},
  \bibinfo{author}{\bibfnamefont{D.~L.} \bibnamefont{Sun}},
  \bibinfo{author}{\bibfnamefont{Y.}~\bibnamefont{Sidis}},
  \bibinfo{author}{\bibfnamefont{A.}~\bibnamefont{Schneidewind}},
  \bibinfo{author}{\bibfnamefont{K.}~\bibnamefont{Hradil}},
  \bibinfo{author}{\bibfnamefont{D.}~\bibnamefont{Haug}},
  \bibinfo{author}{\bibfnamefont{C.~T.} \bibnamefont{Lin}},
  \bibinfo{author}{\bibfnamefont{B.}~\bibnamefont{Keimer}},
  \bibnamefont{et~al.}, \bibinfo{journal}{Nat Phys}
  \textbf{\bibinfo{volume}{6}}, \bibinfo{pages}{178} (\bibinfo{year}{2010}),
  ISSN \bibinfo{issn}{1745-2473},
  \urlprefix\url{http://dx.doi.org/10.1038/nphys1483}.

\bibitem[{\citenamefont{Li et~al.}(2017)\citenamefont{Li, Zhang, Deng, Xu, Mo,
  Yi, Ding, Hashimoto, Moore, Lu et~al.}}]{Li2017}
\bibinfo{author}{\bibfnamefont{W.}~\bibnamefont{Li}},
  \bibinfo{author}{\bibfnamefont{Y.}~\bibnamefont{Zhang}},
  \bibinfo{author}{\bibfnamefont{P.}~\bibnamefont{Deng}},
  \bibinfo{author}{\bibfnamefont{Z.}~\bibnamefont{Xu}},
  \bibinfo{author}{\bibfnamefont{S.-K.} \bibnamefont{Mo}},
  \bibinfo{author}{\bibfnamefont{M.}~\bibnamefont{Yi}},
  \bibinfo{author}{\bibfnamefont{H.}~\bibnamefont{Ding}},
  \bibinfo{author}{\bibfnamefont{M.}~\bibnamefont{Hashimoto}},
  \bibinfo{author}{\bibfnamefont{R.~G.} \bibnamefont{Moore}},
  \bibinfo{author}{\bibfnamefont{D.-H.} \bibnamefont{Lu}},
  \bibnamefont{et~al.}, \bibinfo{journal}{Nature Physics}
  \textbf{\bibinfo{volume}{13}}, \bibinfo{pages}{957} (\bibinfo{year}{2017}),
  \urlprefix\url{https://www.nature.com/articles/nphys4186}.

\bibitem[{\citenamefont{Yim et~al.}(2018)\citenamefont{Yim, Trainer, Aluru,
  Chi, Hardy, Liang, Bonn, and Wahl}}]{Yim2018}
\bibinfo{author}{\bibfnamefont{C.~M.} \bibnamefont{Yim}},
  \bibinfo{author}{\bibfnamefont{C.}~\bibnamefont{Trainer}},
  \bibinfo{author}{\bibfnamefont{R.}~\bibnamefont{Aluru}},
  \bibinfo{author}{\bibfnamefont{S.}~\bibnamefont{Chi}},
  \bibinfo{author}{\bibfnamefont{W.~N.} \bibnamefont{Hardy}},
  \bibinfo{author}{\bibfnamefont{R.}~\bibnamefont{Liang}},
  \bibinfo{author}{\bibfnamefont{D.}~\bibnamefont{Bonn}}, \bibnamefont{and}
  \bibinfo{author}{\bibfnamefont{P.}~\bibnamefont{Wahl}},
  \bibinfo{journal}{Nature Communications} \textbf{\bibinfo{volume}{9}},
  \bibinfo{pages}{2602} (\bibinfo{year}{2018}), ISSN \bibinfo{issn}{2041-1723},
  \urlprefix\url{https://doi.org/10.1038/s41467-018-04909-y}.

\bibitem[{\citenamefont{Lederer et~al.}(2015)\citenamefont{Lederer, Schattner,
  Berg, and Kivelson}}]{Lederer2015}
\bibinfo{author}{\bibfnamefont{S.}~\bibnamefont{Lederer}},
  \bibinfo{author}{\bibfnamefont{Y.}~\bibnamefont{Schattner}},
  \bibinfo{author}{\bibfnamefont{E.}~\bibnamefont{Berg}}, \bibnamefont{and}
  \bibinfo{author}{\bibfnamefont{S.}~\bibnamefont{Kivelson}},
  \bibinfo{journal}{Phys. Rev. Lett.} \textbf{\bibinfo{volume}{114}},
  \bibinfo{pages}{097001} (\bibinfo{year}{2015}),
  \urlprefix\url{https://journals.aps.org/prl/abstract/10.1103/PhysRevLett.114.097001}.

\bibitem[{\citenamefont{Schattner et~al.}(2016)\citenamefont{Schattner,
  Lederer, Kivelson, and Berg}}]{Schattner2016}
\bibinfo{author}{\bibfnamefont{Y.}~\bibnamefont{Schattner}},
  \bibinfo{author}{\bibfnamefont{S.}~\bibnamefont{Lederer}},
  \bibinfo{author}{\bibfnamefont{S.~A.} \bibnamefont{Kivelson}},
  \bibnamefont{and} \bibinfo{author}{\bibfnamefont{E.}~\bibnamefont{Berg}},
  \bibinfo{journal}{Physical Review X} \textbf{\bibinfo{volume}{6}},
  \bibinfo{pages}{031028} (\bibinfo{year}{2016}),
  \urlprefix\url{https://journals.aps.org/prx/pdf/10.1103/PhysRevX.6.031028}.

\bibitem[{\citenamefont{Wang et~al.}(2013{\natexlab{b}})\citenamefont{Wang,
  Kreisel, Zabolotnyy, Borisenko, B\"uchner, Maier, Hirschfeld, and
  Scalapino}}]{Wang2013}
\bibinfo{author}{\bibfnamefont{Y.}~\bibnamefont{Wang}},
  \bibinfo{author}{\bibfnamefont{A.}~\bibnamefont{Kreisel}},
  \bibinfo{author}{\bibfnamefont{V.~B.} \bibnamefont{Zabolotnyy}},
  \bibinfo{author}{\bibfnamefont{S.~V.} \bibnamefont{Borisenko}},
  \bibinfo{author}{\bibfnamefont{B.}~\bibnamefont{B\"uchner}},
  \bibinfo{author}{\bibfnamefont{T.~A.} \bibnamefont{Maier}},
  \bibinfo{author}{\bibfnamefont{P.~J.} \bibnamefont{Hirschfeld}},
  \bibnamefont{and} \bibinfo{author}{\bibfnamefont{D.~J.}
  \bibnamefont{Scalapino}}, \bibinfo{journal}{Phys. Rev. B}
  \textbf{\bibinfo{volume}{88}}, \bibinfo{pages}{174516}
  (\bibinfo{year}{2013}{\natexlab{b}}),
  \urlprefix\url{http://link.aps.org/doi/10.1103/PhysRevB.88.174516}.

\bibitem[{\citenamefont{Hoffman et~al.}(2002)\citenamefont{Hoffman, McElroy,
  Lee, Lang, Eisaki, Uchida, and Davis}}]{Hoffman2002}
\bibinfo{author}{\bibfnamefont{J.~E.} \bibnamefont{Hoffman}},
  \bibinfo{author}{\bibfnamefont{K.}~\bibnamefont{McElroy}},
  \bibinfo{author}{\bibfnamefont{D.-H.} \bibnamefont{Lee}},
  \bibinfo{author}{\bibfnamefont{K.~M.} \bibnamefont{Lang}},
  \bibinfo{author}{\bibfnamefont{H.}~\bibnamefont{Eisaki}},
  \bibinfo{author}{\bibfnamefont{S.}~\bibnamefont{Uchida}}, \bibnamefont{and}
  \bibinfo{author}{\bibfnamefont{J.~C.} \bibnamefont{Davis}},
  \bibinfo{journal}{Science} \textbf{\bibinfo{volume}{297}},
  \bibinfo{pages}{1148} (\bibinfo{year}{2002}),
  \urlprefix\url{http://www.sciencemag.org/content/297/5584/1148.abstract}.

\bibitem[{\citenamefont{Aynajian et~al.}(2012)\citenamefont{Aynajian,
  da~Silva~Neto, Gyenis, Baumbach, Thompson, Fisk, Bauer, and
  Yazdani}}]{Aynajian2012}
\bibinfo{author}{\bibfnamefont{P.}~\bibnamefont{Aynajian}},
  \bibinfo{author}{\bibfnamefont{E.~H.} \bibnamefont{da~Silva~Neto}},
  \bibinfo{author}{\bibfnamefont{A.}~\bibnamefont{Gyenis}},
  \bibinfo{author}{\bibfnamefont{R.~E.} \bibnamefont{Baumbach}},
  \bibinfo{author}{\bibfnamefont{J.~D.} \bibnamefont{Thompson}},
  \bibinfo{author}{\bibfnamefont{Z.}~\bibnamefont{Fisk}},
  \bibinfo{author}{\bibfnamefont{E.~D.} \bibnamefont{Bauer}}, \bibnamefont{and}
  \bibinfo{author}{\bibfnamefont{A.}~\bibnamefont{Yazdani}},
  \bibinfo{journal}{Nature} \textbf{\bibinfo{volume}{486}},
  \bibinfo{pages}{201} (\bibinfo{year}{2012}), ISSN \bibinfo{issn}{0028-0836},
  \urlprefix\url{http://dx.doi.org/10.1038/nature11204}.

\bibitem[{\citenamefont{Allan et~al.}(2012)\citenamefont{Allan, Rost,
  Mackenzie, Xie, Davis, Kihou, Lee, Iyo, Eisaki, and Chuang}}]{Allan2012}
\bibinfo{author}{\bibfnamefont{M.~P.} \bibnamefont{Allan}},
  \bibinfo{author}{\bibfnamefont{A.~W.} \bibnamefont{Rost}},
  \bibinfo{author}{\bibfnamefont{A.~P.} \bibnamefont{Mackenzie}},
  \bibinfo{author}{\bibfnamefont{Y.}~\bibnamefont{Xie}},
  \bibinfo{author}{\bibfnamefont{J.~C.} \bibnamefont{Davis}},
  \bibinfo{author}{\bibfnamefont{K.}~\bibnamefont{Kihou}},
  \bibinfo{author}{\bibfnamefont{C.~H.} \bibnamefont{Lee}},
  \bibinfo{author}{\bibfnamefont{A.}~\bibnamefont{Iyo}},
  \bibinfo{author}{\bibfnamefont{H.}~\bibnamefont{Eisaki}}, \bibnamefont{and}
  \bibinfo{author}{\bibfnamefont{T.-M.} \bibnamefont{Chuang}},
  \bibinfo{journal}{Science} \textbf{\bibinfo{volume}{336}},
  \bibinfo{pages}{563} (\bibinfo{year}{2012}),
  \urlprefix\url{http://www.sciencemag.org/content/336/6081/563.abstract}.

\bibitem[{\citenamefont{H\"anke et~al.}(2012)\citenamefont{H\"anke, Sykora,
  Schlegel, Baumann, Harnagea, Wurmehl, Daghofer, B\"uchner, van~den Brink, and
  Hess}}]{Haenke2012}
\bibinfo{author}{\bibfnamefont{T.}~\bibnamefont{H\"anke}},
  \bibinfo{author}{\bibfnamefont{S.}~\bibnamefont{Sykora}},
  \bibinfo{author}{\bibfnamefont{R.}~\bibnamefont{Schlegel}},
  \bibinfo{author}{\bibfnamefont{D.}~\bibnamefont{Baumann}},
  \bibinfo{author}{\bibfnamefont{L.}~\bibnamefont{Harnagea}},
  \bibinfo{author}{\bibfnamefont{S.}~\bibnamefont{Wurmehl}},
  \bibinfo{author}{\bibfnamefont{M.}~\bibnamefont{Daghofer}},
  \bibinfo{author}{\bibfnamefont{B.}~\bibnamefont{B\"uchner}},
  \bibinfo{author}{\bibfnamefont{J.}~\bibnamefont{van~den Brink}},
  \bibnamefont{and} \bibinfo{author}{\bibfnamefont{C.}~\bibnamefont{Hess}},
  \bibinfo{journal}{Phys. Rev. Lett.} \textbf{\bibinfo{volume}{108}},
  \bibinfo{pages}{127001} (\bibinfo{year}{2012}),
  \urlprefix\url{http://link.aps.org/doi/10.1103/PhysRevLett.108.127001}.

\bibitem[{\citenamefont{Hess et~al.}(2013)\citenamefont{Hess, Sykora, H\"anke,
  Schlegel, Baumann, Zabolotnyy, Harnagea, Wurmehl, van~den Brink, and
  B\"uchner}}]{Hess2013}
\bibinfo{author}{\bibfnamefont{C.}~\bibnamefont{Hess}},
  \bibinfo{author}{\bibfnamefont{S.}~\bibnamefont{Sykora}},
  \bibinfo{author}{\bibfnamefont{T.}~\bibnamefont{H\"anke}},
  \bibinfo{author}{\bibfnamefont{R.}~\bibnamefont{Schlegel}},
  \bibinfo{author}{\bibfnamefont{D.}~\bibnamefont{Baumann}},
  \bibinfo{author}{\bibfnamefont{V.~B.} \bibnamefont{Zabolotnyy}},
  \bibinfo{author}{\bibfnamefont{L.}~\bibnamefont{Harnagea}},
  \bibinfo{author}{\bibfnamefont{S.}~\bibnamefont{Wurmehl}},
  \bibinfo{author}{\bibfnamefont{J.}~\bibnamefont{van~den Brink}},
  \bibnamefont{and}
  \bibinfo{author}{\bibfnamefont{B.}~\bibnamefont{B\"uchner}},
  \bibinfo{journal}{Phys. Rev. Lett.} \textbf{\bibinfo{volume}{110}},
  \bibinfo{pages}{017006} (\bibinfo{year}{2013}),
  \urlprefix\url{http://link.aps.org/doi/10.1103/PhysRevLett.110.017006}.

\bibitem[{\citenamefont{Grothe et~al.}(2013)\citenamefont{Grothe, Johnston,
  Chi, Dosanjh, Burke, and Pennec}}]{Grothe2013}
\bibinfo{author}{\bibfnamefont{S.}~\bibnamefont{Grothe}},
  \bibinfo{author}{\bibfnamefont{S.}~\bibnamefont{Johnston}},
  \bibinfo{author}{\bibfnamefont{S.}~\bibnamefont{Chi}},
  \bibinfo{author}{\bibfnamefont{P.}~\bibnamefont{Dosanjh}},
  \bibinfo{author}{\bibfnamefont{S.~A.} \bibnamefont{Burke}}, \bibnamefont{and}
  \bibinfo{author}{\bibfnamefont{Y.}~\bibnamefont{Pennec}},
  \bibinfo{journal}{Phys. Rev. Lett.} \textbf{\bibinfo{volume}{111}},
  \bibinfo{pages}{246804} (\bibinfo{year}{2013}),
  \urlprefix\url{https://link.aps.org/doi/10.1103/PhysRevLett.111.246804}.

\bibitem[{\citenamefont{Allan et~al.}(2015)\citenamefont{Allan, Lee, Rost,
  Fischer, Massee, Kihou, Lee, Iyo, Eisaki, Chuang et~al.}}]{Allan2015}
\bibinfo{author}{\bibfnamefont{M.~P.} \bibnamefont{Allan}},
  \bibinfo{author}{\bibfnamefont{K.}~\bibnamefont{Lee}},
  \bibinfo{author}{\bibfnamefont{A.~W.} \bibnamefont{Rost}},
  \bibinfo{author}{\bibfnamefont{M.~H.} \bibnamefont{Fischer}},
  \bibinfo{author}{\bibfnamefont{F.}~\bibnamefont{Massee}},
  \bibinfo{author}{\bibfnamefont{K.}~\bibnamefont{Kihou}},
  \bibinfo{author}{\bibfnamefont{C.-H.} \bibnamefont{Lee}},
  \bibinfo{author}{\bibfnamefont{A.}~\bibnamefont{Iyo}},
  \bibinfo{author}{\bibfnamefont{H.}~\bibnamefont{Eisaki}},
  \bibinfo{author}{\bibfnamefont{T.-M.} \bibnamefont{Chuang}},
  \bibnamefont{et~al.}, \bibinfo{journal}{Nat Phys}
  \textbf{\bibinfo{volume}{11}}, \bibinfo{pages}{177} (\bibinfo{year}{2015}),
  ISSN \bibinfo{issn}{1745-2473},
  \urlprefix\url{http://dx.doi.org/10.1038/nphys3187}.

\bibitem[{\citenamefont{Wang et~al.}(2017)\citenamefont{Wang, Walkup, Derry,
  Scaffidi, Rak, Vig, Kogar, Zeljkovic, Husain, Santos et~al.}}]{Wang2017}
\bibinfo{author}{\bibfnamefont{Z.}~\bibnamefont{Wang}},
  \bibinfo{author}{\bibfnamefont{D.}~\bibnamefont{Walkup}},
  \bibinfo{author}{\bibfnamefont{P.}~\bibnamefont{Derry}},
  \bibinfo{author}{\bibfnamefont{T.}~\bibnamefont{Scaffidi}},
  \bibinfo{author}{\bibfnamefont{M.}~\bibnamefont{Rak}},
  \bibinfo{author}{\bibfnamefont{S.}~\bibnamefont{Vig}},
  \bibinfo{author}{\bibfnamefont{A.}~\bibnamefont{Kogar}},
  \bibinfo{author}{\bibfnamefont{I.}~\bibnamefont{Zeljkovic}},
  \bibinfo{author}{\bibfnamefont{A.}~\bibnamefont{Husain}},
  \bibinfo{author}{\bibfnamefont{L.~H.} \bibnamefont{Santos}},
  \bibnamefont{et~al.}, \bibinfo{journal}{Nat Phys}
  \textbf{\bibinfo{volume}{13}}, \bibinfo{pages}{799} (\bibinfo{year}{2017}),
  ISSN \bibinfo{issn}{1745-2473},
  \urlprefix\url{http://dx.doi.org/10.1038/nphys4107}.

\bibitem[{\citenamefont{Borisenko et~al.}(2010)\citenamefont{Borisenko,
  Zabolotnyy, Evtushinsky, Kim, Morozov, Yaresko, Kordyuk, Behr, Vasiliev,
  Follath et~al.}}]{Borisenko2010}
\bibinfo{author}{\bibfnamefont{S.~V.} \bibnamefont{Borisenko}},
  \bibinfo{author}{\bibfnamefont{V.~B.} \bibnamefont{Zabolotnyy}},
  \bibinfo{author}{\bibfnamefont{D.~V.} \bibnamefont{Evtushinsky}},
  \bibinfo{author}{\bibfnamefont{T.~K.} \bibnamefont{Kim}},
  \bibinfo{author}{\bibfnamefont{I.~V.} \bibnamefont{Morozov}},
  \bibinfo{author}{\bibfnamefont{A.~N.} \bibnamefont{Yaresko}},
  \bibinfo{author}{\bibfnamefont{A.~A.} \bibnamefont{Kordyuk}},
  \bibinfo{author}{\bibfnamefont{G.}~\bibnamefont{Behr}},
  \bibinfo{author}{\bibfnamefont{A.}~\bibnamefont{Vasiliev}},
  \bibinfo{author}{\bibfnamefont{R.}~\bibnamefont{Follath}},
  \bibnamefont{et~al.}, \bibinfo{journal}{Phys. Rev. Lett.}
  \textbf{\bibinfo{volume}{105}}, \bibinfo{pages}{067002}
  (\bibinfo{year}{2010}),
  \urlprefix\url{http://link.aps.org/doi/10.1103/PhysRevLett.105.067002}.

\bibitem[{\citenamefont{Zeng et~al.}(2013)\citenamefont{Zeng, Watanabe, Zhang,
  Li, Besara, Siegrist, Xing, Wang, Jin, Goswami et~al.}}]{Zeng2013}
\bibinfo{author}{\bibfnamefont{B.}~\bibnamefont{Zeng}},
  \bibinfo{author}{\bibfnamefont{D.}~\bibnamefont{Watanabe}},
  \bibinfo{author}{\bibfnamefont{Q.~R.} \bibnamefont{Zhang}},
  \bibinfo{author}{\bibfnamefont{G.}~\bibnamefont{Li}},
  \bibinfo{author}{\bibfnamefont{T.}~\bibnamefont{Besara}},
  \bibinfo{author}{\bibfnamefont{T.}~\bibnamefont{Siegrist}},
  \bibinfo{author}{\bibfnamefont{L.~Y.} \bibnamefont{Xing}},
  \bibinfo{author}{\bibfnamefont{X.~C.} \bibnamefont{Wang}},
  \bibinfo{author}{\bibfnamefont{C.~Q.} \bibnamefont{Jin}},
  \bibinfo{author}{\bibfnamefont{P.}~\bibnamefont{Goswami}},
  \bibnamefont{et~al.}, \bibinfo{journal}{Phys. Rev. B}
  \textbf{\bibinfo{volume}{88}}, \bibinfo{pages}{144518}
  (\bibinfo{year}{2013}),
  \urlprefix\url{http://link.aps.org/doi/10.1103/PhysRevB.88.144518}.

\bibitem[{\citenamefont{Aswartham et~al.}(2011)\citenamefont{Aswartham, Behr,
  Harnagea, Bombor, Bachmann, Morozov, Zabolotnyy, Kordyuk, Kim, Evtushinsky
  et~al.}}]{Aswartham2011a}
\bibinfo{author}{\bibfnamefont{S.}~\bibnamefont{Aswartham}},
  \bibinfo{author}{\bibfnamefont{G.}~\bibnamefont{Behr}},
  \bibinfo{author}{\bibfnamefont{L.}~\bibnamefont{Harnagea}},
  \bibinfo{author}{\bibfnamefont{D.}~\bibnamefont{Bombor}},
  \bibinfo{author}{\bibfnamefont{A.}~\bibnamefont{Bachmann}},
  \bibinfo{author}{\bibfnamefont{I.~V.} \bibnamefont{Morozov}},
  \bibinfo{author}{\bibfnamefont{V.~B.} \bibnamefont{Zabolotnyy}},
  \bibinfo{author}{\bibfnamefont{A.~A.} \bibnamefont{Kordyuk}},
  \bibinfo{author}{\bibfnamefont{T.~K.} \bibnamefont{Kim}},
  \bibinfo{author}{\bibfnamefont{D.~V.} \bibnamefont{Evtushinsky}},
  \bibnamefont{et~al.}, \bibinfo{journal}{Phys. Rev. B}
  \textbf{\bibinfo{volume}{84}}, \bibinfo{pages}{054534}
  (\bibinfo{year}{2011}),
  \urlprefix\url{http://link.aps.org/doi/10.1103/PhysRevB.84.054534}.

\bibitem[{\citenamefont{Pitcher et~al.}(2010)\citenamefont{Pitcher, Lancaster,
  Wright, Franke, Steele, Baker, Pratt, Thomas, Parker, Blundell
  et~al.}}]{Pitcher2010}
\bibinfo{author}{\bibfnamefont{M.~J.} \bibnamefont{Pitcher}},
  \bibinfo{author}{\bibfnamefont{T.}~\bibnamefont{Lancaster}},
  \bibinfo{author}{\bibfnamefont{J.~D.} \bibnamefont{Wright}},
  \bibinfo{author}{\bibfnamefont{I.}~\bibnamefont{Franke}},
  \bibinfo{author}{\bibfnamefont{A.~J.} \bibnamefont{Steele}},
  \bibinfo{author}{\bibfnamefont{P.~J.} \bibnamefont{Baker}},
  \bibinfo{author}{\bibfnamefont{F.~L.} \bibnamefont{Pratt}},
  \bibinfo{author}{\bibfnamefont{W.~T.} \bibnamefont{Thomas}},
  \bibinfo{author}{\bibfnamefont{D.~R.} \bibnamefont{Parker}},
  \bibinfo{author}{\bibfnamefont{S.~J.} \bibnamefont{Blundell}},
  \bibnamefont{et~al.}, \bibinfo{journal}{J. Am. Chem. Soc.}
  \textbf{\bibinfo{volume}{132}}, \bibinfo{pages}{10467}
  (\bibinfo{year}{2010}),
  \urlprefix\url{http://pubs.acs.org/doi/abs/10.1021/ja103196c}.

\bibitem[{\citenamefont{Qureshi et~al.}(2012)\citenamefont{Qureshi, Steffens,
  Drees, Komarek, Lamago, Sidis, Harnagea, Grafe, Wurmehl, B\"uchner
  et~al.}}]{Qureshi2012}
\bibinfo{author}{\bibfnamefont{N.}~\bibnamefont{Qureshi}},
  \bibinfo{author}{\bibfnamefont{P.}~\bibnamefont{Steffens}},
  \bibinfo{author}{\bibfnamefont{Y.}~\bibnamefont{Drees}},
  \bibinfo{author}{\bibfnamefont{A.~C.} \bibnamefont{Komarek}},
  \bibinfo{author}{\bibfnamefont{D.}~\bibnamefont{Lamago}},
  \bibinfo{author}{\bibfnamefont{Y.}~\bibnamefont{Sidis}},
  \bibinfo{author}{\bibfnamefont{L.}~\bibnamefont{Harnagea}},
  \bibinfo{author}{\bibfnamefont{H.-J.} \bibnamefont{Grafe}},
  \bibinfo{author}{\bibfnamefont{S.}~\bibnamefont{Wurmehl}},
  \bibinfo{author}{\bibfnamefont{B.}~\bibnamefont{B\"uchner}},
  \bibnamefont{et~al.}, \bibinfo{journal}{Phys. Rev. Lett.}
  \textbf{\bibinfo{volume}{108}}, \bibinfo{pages}{117001}
  (\bibinfo{year}{2012}),
  \urlprefix\url{http://link.aps.org/doi/10.1103/PhysRevLett.108.117001}.

\bibitem[{\citenamefont{Knolle et~al.}(2012)\citenamefont{Knolle, Zabolotnyy,
  Eremin, Borisenko, Qureshi, Braden, Evtushinsky, Kim, Kordyuk, Sykora
  et~al.}}]{Knolle2012}
\bibinfo{author}{\bibfnamefont{J.}~\bibnamefont{Knolle}},
  \bibinfo{author}{\bibfnamefont{V.~B.} \bibnamefont{Zabolotnyy}},
  \bibinfo{author}{\bibfnamefont{I.}~\bibnamefont{Eremin}},
  \bibinfo{author}{\bibfnamefont{S.~V.} \bibnamefont{Borisenko}},
  \bibinfo{author}{\bibfnamefont{N.}~\bibnamefont{Qureshi}},
  \bibinfo{author}{\bibfnamefont{M.}~\bibnamefont{Braden}},
  \bibinfo{author}{\bibfnamefont{D.~V.} \bibnamefont{Evtushinsky}},
  \bibinfo{author}{\bibfnamefont{T.~K.} \bibnamefont{Kim}},
  \bibinfo{author}{\bibfnamefont{A.~A.} \bibnamefont{Kordyuk}},
  \bibinfo{author}{\bibfnamefont{S.}~\bibnamefont{Sykora}},
  \bibnamefont{et~al.}, \bibinfo{journal}{Phys. Rev. B}
  \textbf{\bibinfo{volume}{86}}, \bibinfo{pages}{174519}
  (\bibinfo{year}{2012}),
  \urlprefix\url{http://link.aps.org/doi/10.1103/PhysRevB.86.174519}.

\bibitem[{\citenamefont{Ahn et~al.}(2014)\citenamefont{Ahn, Eremin, Knolle,
  Zabolotnyy, Borisenko, B\"uchner, and Chubukov}}]{Ahn2014a}
\bibinfo{author}{\bibfnamefont{F.}~\bibnamefont{Ahn}},
  \bibinfo{author}{\bibfnamefont{I.}~\bibnamefont{Eremin}},
  \bibinfo{author}{\bibfnamefont{J.}~\bibnamefont{Knolle}},
  \bibinfo{author}{\bibfnamefont{V.~B.} \bibnamefont{Zabolotnyy}},
  \bibinfo{author}{\bibfnamefont{S.~V.} \bibnamefont{Borisenko}},
  \bibinfo{author}{\bibfnamefont{B.}~\bibnamefont{B\"uchner}},
  \bibnamefont{and} \bibinfo{author}{\bibfnamefont{A.~V.}
  \bibnamefont{Chubukov}}, \bibinfo{journal}{Phys. Rev. B}
  \textbf{\bibinfo{volume}{89}}, \bibinfo{pages}{144513}
  (\bibinfo{year}{2014}),
  \urlprefix\url{http://link.aps.org/doi/10.1103/PhysRevB.89.144513}.

\bibitem[{\citenamefont{Borisenko et~al.}(2012)\citenamefont{Borisenko,
  Zabolotnyy, Kordyuk, Evtushinsky, Kim, Morozov, Follath, and
  B\"{u}chner}}]{Borisenko2012}
\bibinfo{author}{\bibfnamefont{S.~V.} \bibnamefont{Borisenko}},
  \bibinfo{author}{\bibfnamefont{V.~B.} \bibnamefont{Zabolotnyy}},
  \bibinfo{author}{\bibfnamefont{A.~A.} \bibnamefont{Kordyuk}},
  \bibinfo{author}{\bibfnamefont{D.~V.} \bibnamefont{Evtushinsky}},
  \bibinfo{author}{\bibfnamefont{T.~K.} \bibnamefont{Kim}},
  \bibinfo{author}{\bibfnamefont{I.~V.} \bibnamefont{Morozov}},
  \bibinfo{author}{\bibfnamefont{R.}~\bibnamefont{Follath}}, \bibnamefont{and}
  \bibinfo{author}{\bibfnamefont{B.}~\bibnamefont{B\"{u}chner}},
  \bibinfo{journal}{Symmetry} \textbf{\bibinfo{volume}{4}},
  \bibinfo{pages}{251} (\bibinfo{year}{2012}),
  \urlprefix\url{http://www.mdpi.com/2073-8994/4/1/251}.

\bibitem[{\citenamefont{Chi et~al.}(2014)\citenamefont{Chi, Johnston, Levy,
  Grothe, Szedlak, Ludbrook, Liang, Dosanjh, Burke, Damascelli
  et~al.}}]{Chi2014}
\bibinfo{author}{\bibfnamefont{S.}~\bibnamefont{Chi}},
  \bibinfo{author}{\bibfnamefont{S.}~\bibnamefont{Johnston}},
  \bibinfo{author}{\bibfnamefont{G.}~\bibnamefont{Levy}},
  \bibinfo{author}{\bibfnamefont{S.}~\bibnamefont{Grothe}},
  \bibinfo{author}{\bibfnamefont{R.}~\bibnamefont{Szedlak}},
  \bibinfo{author}{\bibfnamefont{B.}~\bibnamefont{Ludbrook}},
  \bibinfo{author}{\bibfnamefont{R.}~\bibnamefont{Liang}},
  \bibinfo{author}{\bibfnamefont{P.}~\bibnamefont{Dosanjh}},
  \bibinfo{author}{\bibfnamefont{S.~A.} \bibnamefont{Burke}},
  \bibinfo{author}{\bibfnamefont{A.}~\bibnamefont{Damascelli}},
  \bibnamefont{et~al.}, \bibinfo{journal}{Phys. Rev. B}
  \textbf{\bibinfo{volume}{89}}, \bibinfo{pages}{104522}
  (\bibinfo{year}{2014}),
  \urlprefix\url{http://link.aps.org/doi/10.1103/PhysRevB.89.104522}.

\bibitem[{\citenamefont{Grothe et~al.}(2012)\citenamefont{Grothe, Chi, Dosanjh,
  Liang, Hardy, Burke, Bonn, and Pennec}}]{Grothe2012}
\bibinfo{author}{\bibfnamefont{S.}~\bibnamefont{Grothe}},
  \bibinfo{author}{\bibfnamefont{S.}~\bibnamefont{Chi}},
  \bibinfo{author}{\bibfnamefont{P.}~\bibnamefont{Dosanjh}},
  \bibinfo{author}{\bibfnamefont{R.}~\bibnamefont{Liang}},
  \bibinfo{author}{\bibfnamefont{W.~N.} \bibnamefont{Hardy}},
  \bibinfo{author}{\bibfnamefont{S.~A.} \bibnamefont{Burke}},
  \bibinfo{author}{\bibfnamefont{D.~A.} \bibnamefont{Bonn}}, \bibnamefont{and}
  \bibinfo{author}{\bibfnamefont{Y.}~\bibnamefont{Pennec}},
  \bibinfo{journal}{Phys. Rev. B} \textbf{\bibinfo{volume}{86}},
  \bibinfo{pages}{174503} (\bibinfo{year}{2012}),
  \urlprefix\url{http://link.aps.org/doi/10.1103/PhysRevB.86.174503}.

\bibitem[{\citenamefont{Schlegel et~al.}(2017)\citenamefont{Schlegel, Nag,
  Baumann, Beck, Wurmehl, B\"uchner, and Hess}}]{Schlegel2017}
\bibinfo{author}{\bibfnamefont{R.}~\bibnamefont{Schlegel}},
  \bibinfo{author}{\bibfnamefont{P.~K.} \bibnamefont{Nag}},
  \bibinfo{author}{\bibfnamefont{D.}~\bibnamefont{Baumann}},
  \bibinfo{author}{\bibfnamefont{R.}~\bibnamefont{Beck}},
  \bibinfo{author}{\bibfnamefont{S.}~\bibnamefont{Wurmehl}},
  \bibinfo{author}{\bibfnamefont{B.}~\bibnamefont{B\"uchner}},
  \bibnamefont{and} \bibinfo{author}{\bibfnamefont{C.}~\bibnamefont{Hess}},
  \bibinfo{journal}{Physica Status Solidi B} \textbf{\bibinfo{volume}{254}},
  \bibinfo{pages}{1600159} (\bibinfo{year}{2017}), ISSN
  \bibinfo{issn}{1521-3951}, \bibinfo{note}{1600159},
  \urlprefix\url{http://dx.doi.org/10.1002/pssb.201600159}.

\bibitem[{SM()}]{SM}
\bibinfo{title}{{See Supplemental Material at [URL] for an overview of
  the full data set (movie).}}

\bibitem[{\citenamefont{Chi et~al.}(2017)\citenamefont{Chi, Aluru, Grothe,
  Kreisel, Singh, Andersen, Hardy, Liang, Bonn, Burke et~al.}}]{Chi2017}
\bibinfo{author}{\bibfnamefont{S.}~\bibnamefont{Chi}},
  \bibinfo{author}{\bibfnamefont{R.}~\bibnamefont{Aluru}},
  \bibinfo{author}{\bibfnamefont{S.}~\bibnamefont{Grothe}},
  \bibinfo{author}{\bibfnamefont{A.}~\bibnamefont{Kreisel}},
  \bibinfo{author}{\bibfnamefont{U.~R.} \bibnamefont{Singh}},
  \bibinfo{author}{\bibfnamefont{B.~M.} \bibnamefont{Andersen}},
  \bibinfo{author}{\bibfnamefont{W.~N.} \bibnamefont{Hardy}},
  \bibinfo{author}{\bibfnamefont{R.}~\bibnamefont{Liang}},
  \bibinfo{author}{\bibfnamefont{D.~A.} \bibnamefont{Bonn}},
  \bibinfo{author}{\bibfnamefont{S.~A.} \bibnamefont{Burke}},
  \bibnamefont{et~al.}, \bibinfo{journal}{Nature Comm.}
  \textbf{\bibinfo{volume}{8}}, \bibinfo{pages}{15996} (\bibinfo{year}{2017}),
  \urlprefix\url{http://dx.doi.org/10.1038/ncomms15996}.

\bibitem[{\citenamefont{Balatsky et~al.}(2006)\citenamefont{Balatsky, Vekhter,
  and Zhu}}]{Balatsky2006}
\bibinfo{author}{\bibfnamefont{A.~V.} \bibnamefont{Balatsky}},
  \bibinfo{author}{\bibfnamefont{I.}~\bibnamefont{Vekhter}}, \bibnamefont{and}
  \bibinfo{author}{\bibfnamefont{J.-X.} \bibnamefont{Zhu}},
  \bibinfo{journal}{Rev. Mod. Phys.} \textbf{\bibinfo{volume}{78}},
  \bibinfo{pages}{373} (\bibinfo{year}{2006}),
  \urlprefix\url{http://link.aps.org/doi/10.1103/RevModPhys.78.373}.

\bibitem[{\citenamefont{Borisenko et~al.}(2016)\citenamefont{Borisenko,
  Evtushinsky, Liu, Morozov, Kappenberger, Wurmehl, B\"uchner, Yaresko, Kim,
  Hoesch et~al.}}]{Borisenko2016}
\bibinfo{author}{\bibfnamefont{S.~V.} \bibnamefont{Borisenko}},
  \bibinfo{author}{\bibfnamefont{D.~V.} \bibnamefont{Evtushinsky}},
  \bibinfo{author}{\bibfnamefont{Z.-H.} \bibnamefont{Liu}},
  \bibinfo{author}{\bibfnamefont{I.}~\bibnamefont{Morozov}},
  \bibinfo{author}{\bibfnamefont{R.}~\bibnamefont{Kappenberger}},
  \bibinfo{author}{\bibfnamefont{S.}~\bibnamefont{Wurmehl}},
  \bibinfo{author}{\bibfnamefont{B.}~\bibnamefont{B\"uchner}},
  \bibinfo{author}{\bibfnamefont{A.~N.} \bibnamefont{Yaresko}},
  \bibinfo{author}{\bibfnamefont{T.~K.} \bibnamefont{Kim}},
  \bibinfo{author}{\bibfnamefont{M.}~\bibnamefont{Hoesch}},
  \bibnamefont{et~al.}, \bibinfo{journal}{Nat Phys}
  \textbf{\bibinfo{volume}{12}}, \bibinfo{pages}{311} (\bibinfo{year}{2016}),
  ISSN \bibinfo{issn}{1745-2473},
  \urlprefix\url{http://dx.doi.org/10.1038/nphys3594}.

\bibitem[{err()}]{error}
\bibinfo{title}{{We estimate the error for ${\Omega}$ from the width of
  the peak at ${\Delta}_1+{\Omega}$ in {Fig}.~3{f} and in {Fig}.~4{e}.}}

\bibitem[{\citenamefont{Hwang et~al.}(2015)\citenamefont{Hwang, Carbotte, Min,
  Kwon, and Timusk}}]{Hwang2015}
\bibinfo{author}{\bibfnamefont{J.}~\bibnamefont{Hwang}},
  \bibinfo{author}{\bibfnamefont{J.~P.} \bibnamefont{Carbotte}},
  \bibinfo{author}{\bibfnamefont{B.~H.} \bibnamefont{Min}},
  \bibinfo{author}{\bibfnamefont{Y.~S.} \bibnamefont{Kwon}}, \bibnamefont{and}
  \bibinfo{author}{\bibfnamefont{T.}~\bibnamefont{Timusk}},
  \bibinfo{journal}{Journal of Physics: Condensed Matter}
  \textbf{\bibinfo{volume}{27}}, \bibinfo{pages}{055701}
  (\bibinfo{year}{2015}),
  \urlprefix\url{http://stacks.iop.org/0953-8984/27/i=5/a=055701}.

\bibitem[{\citenamefont{Brydon et~al.}(2011)\citenamefont{Brydon, Daghofer,
  Timm, and van~den Brink}}]{Brydon2011}
\bibinfo{author}{\bibfnamefont{P.~M.~R.} \bibnamefont{Brydon}},
  \bibinfo{author}{\bibfnamefont{M.}~\bibnamefont{Daghofer}},
  \bibinfo{author}{\bibfnamefont{C.}~\bibnamefont{Timm}}, \bibnamefont{and}
  \bibinfo{author}{\bibfnamefont{J.}~\bibnamefont{van~den Brink}},
  \bibinfo{journal}{Phys. Rev. B} \textbf{\bibinfo{volume}{83}},
  \bibinfo{pages}{060501} (\bibinfo{year}{2011}),
  \urlprefix\url{http://link.aps.org/doi/10.1103/PhysRevB.83.060501}.

\bibitem[{\citenamefont{Yin et~al.}(2014)\citenamefont{Yin, Haule, and
  Kotliar}}]{Yin2014}
\bibinfo{author}{\bibfnamefont{Z.~P.} \bibnamefont{Yin}},
  \bibinfo{author}{\bibfnamefont{K.}~\bibnamefont{Haule}}, \bibnamefont{and}
  \bibinfo{author}{\bibfnamefont{G.}~\bibnamefont{Kotliar}},
  \bibinfo{journal}{Nat Phys} \textbf{\bibinfo{volume}{10}},
  \bibinfo{pages}{845} (\bibinfo{year}{2014}), ISSN \bibinfo{issn}{1745-2473},
  \urlprefix\url{http://dx.doi.org/10.1038/nphys3116}.

\bibitem[{\citenamefont{Saito et~al.}(2015)\citenamefont{Saito, Yamakawa,
  Onari, and Kontani}}]{Saito2015}
\bibinfo{author}{\bibfnamefont{T.}~\bibnamefont{Saito}},
  \bibinfo{author}{\bibfnamefont{Y.}~\bibnamefont{Yamakawa}},
  \bibinfo{author}{\bibfnamefont{S.}~\bibnamefont{Onari}}, \bibnamefont{and}
  \bibinfo{author}{\bibfnamefont{H.}~\bibnamefont{Kontani}},
  \bibinfo{journal}{Phys. Rev. B} \textbf{\bibinfo{volume}{92}},
  \bibinfo{pages}{134522} (\bibinfo{year}{2015}),
  \urlprefix\url{https://link.aps.org/doi/10.1103/PhysRevB.92.134522}.

\bibitem[{\citenamefont{Becker et~al.}(2002)\citenamefont{Becker, H\"ubsch, and
  Sommer}}]{Becker2002}
\bibinfo{author}{\bibfnamefont{K.~W.} \bibnamefont{Becker}},
  \bibinfo{author}{\bibfnamefont{A.}~\bibnamefont{H\"ubsch}}, \bibnamefont{and}
  \bibinfo{author}{\bibfnamefont{T.}~\bibnamefont{Sommer}},
  \bibinfo{journal}{Phys. Rev. B} \textbf{\bibinfo{volume}{66}},
  \bibinfo{pages}{235115} (\bibinfo{year}{2002}),
  \urlprefix\url{https://link.aps.org/doi/10.1103/PhysRevB.66.235115}.

\bibitem[{\citenamefont{Sykora et~al.}(2005)\citenamefont{Sykora, H\"ubsch,
  Becker, Wellein, and Fehske}}]{Sykora2005}
\bibinfo{author}{\bibfnamefont{S.}~\bibnamefont{Sykora}},
  \bibinfo{author}{\bibfnamefont{A.}~\bibnamefont{H\"ubsch}},
  \bibinfo{author}{\bibfnamefont{K.~W.} \bibnamefont{Becker}},
  \bibinfo{author}{\bibfnamefont{G.}~\bibnamefont{Wellein}}, \bibnamefont{and}
  \bibinfo{author}{\bibfnamefont{H.}~\bibnamefont{Fehske}},
  \bibinfo{journal}{Phys. Rev. B} \textbf{\bibinfo{volume}{71}},
  \bibinfo{pages}{045112} (\bibinfo{year}{2005}),
  \urlprefix\url{https://link.aps.org/doi/10.1103/PhysRevB.71.045112}.

\bibitem[{\citenamefont{Cho et~al.}(2016)\citenamefont{Cho, Brink, Fehske,
  Becker, and Sykora}}]{Cho2016}
\bibinfo{author}{\bibfnamefont{D.-N.} \bibnamefont{Cho}},
  \bibinfo{author}{\bibfnamefont{J.~v.~d.} \bibnamefont{Brink}},
  \bibinfo{author}{\bibfnamefont{H.}~\bibnamefont{Fehske}},
  \bibinfo{author}{\bibfnamefont{K.~W.} \bibnamefont{Becker}},
  \bibnamefont{and} \bibinfo{author}{\bibfnamefont{S.}~\bibnamefont{Sykora}},
  \textbf{\bibinfo{volume}{6}}, \bibinfo{pages}{22548} (\bibinfo{year}{2016}),
  \urlprefix\url{http://dx.doi.org/10.1038/srep22548}.

\bibitem[{\citenamefont{Polkovnikov et~al.}(2002)\citenamefont{Polkovnikov,
  Vojta, and Sachdev}}]{Polkovnikov2002}
\bibinfo{author}{\bibfnamefont{A.}~\bibnamefont{Polkovnikov}},
  \bibinfo{author}{\bibfnamefont{M.}~\bibnamefont{Vojta}}, \bibnamefont{and}
  \bibinfo{author}{\bibfnamefont{S.}~\bibnamefont{Sachdev}},
  \bibinfo{journal}{Phys. Rev. B} \textbf{\bibinfo{volume}{65}},
  \bibinfo{pages}{220509} (\bibinfo{year}{2002}),
  \urlprefix\url{https://link.aps.org/doi/10.1103/PhysRevB.65.220509}.

\bibitem[{\citenamefont{Rossi and Morr}(2010)}]{Rossi2010}
\bibinfo{author}{\bibfnamefont{E.}~\bibnamefont{Rossi}} \bibnamefont{and}
  \bibinfo{author}{\bibfnamefont{D.~K.} \bibnamefont{Morr}},
  \bibinfo{journal}{Phys. Rev. B} \textbf{\bibinfo{volume}{81}},
  \bibinfo{pages}{054443} (\bibinfo{year}{2010}),
  \urlprefix\url{https://link.aps.org/doi/10.1103/PhysRevB.81.054443}.

\bibitem[{\citenamefont{Hor\'a\ifmmode~\check{c}\else \v{c}\fi{}ek and
  Sasakawa}(1983)}]{Horacek1983}
\bibinfo{author}{\bibfnamefont{J.}~\bibnamefont{Hor\'a\ifmmode~\check{c}\else
  \v{c}\fi{}ek}} \bibnamefont{and}
  \bibinfo{author}{\bibfnamefont{T.}~\bibnamefont{Sasakawa}},
  \bibinfo{journal}{Phys. Rev. A} \textbf{\bibinfo{volume}{28}},
  \bibinfo{pages}{2151} (\bibinfo{year}{1983}),
  \urlprefix\url{https://link.aps.org/doi/10.1103/PhysRevA.28.2151}.

\bibitem[{\citenamefont{Morozov et~al.}(2010)\citenamefont{Morozov, Boltalin,
  Volkova, Vasiliev, Kataeva, Stockert, Abdel-Hafiez, Bombor, Bachmann,
  Harnagea et~al.}}]{Morozov2010}
\bibinfo{author}{\bibfnamefont{I.}~\bibnamefont{Morozov}},
  \bibinfo{author}{\bibfnamefont{A.}~\bibnamefont{Boltalin}},
  \bibinfo{author}{\bibfnamefont{O.}~\bibnamefont{Volkova}},
  \bibinfo{author}{\bibfnamefont{A.}~\bibnamefont{Vasiliev}},
  \bibinfo{author}{\bibfnamefont{O.}~\bibnamefont{Kataeva}},
  \bibinfo{author}{\bibfnamefont{U.}~\bibnamefont{Stockert}},
  \bibinfo{author}{\bibfnamefont{M.}~\bibnamefont{Abdel-Hafiez}},
  \bibinfo{author}{\bibfnamefont{D.}~\bibnamefont{Bombor}},
  \bibinfo{author}{\bibfnamefont{A.}~\bibnamefont{Bachmann}},
  \bibinfo{author}{\bibfnamefont{L.}~\bibnamefont{Harnagea}},
  \bibnamefont{et~al.}, \bibinfo{journal}{Crystal Growth \& Design}
  \textbf{\bibinfo{volume}{10}}, \bibinfo{pages}{4428} (\bibinfo{year}{2010}),
  \urlprefix\url{http://dx.doi.org/10.1021/cg1005538}.

\bibitem[{\citenamefont{Schlegel et~al.}(2014)\citenamefont{Schlegel, H\"anke,
  Baumann, Kaiser, Nag, Voigtl\"ander, Lindackers, B\"uchner, and
  Hess}}]{Schlegel2014}
\bibinfo{author}{\bibfnamefont{R.}~\bibnamefont{Schlegel}},
  \bibinfo{author}{\bibfnamefont{T.}~\bibnamefont{H\"anke}},
  \bibinfo{author}{\bibfnamefont{D.}~\bibnamefont{Baumann}},
  \bibinfo{author}{\bibfnamefont{M.}~\bibnamefont{Kaiser}},
  \bibinfo{author}{\bibfnamefont{P.~K.} \bibnamefont{Nag}},
  \bibinfo{author}{\bibfnamefont{R.}~\bibnamefont{Voigtl\"ander}},
  \bibinfo{author}{\bibfnamefont{D.}~\bibnamefont{Lindackers}},
  \bibinfo{author}{\bibfnamefont{B.}~\bibnamefont{B\"uchner}},
  \bibnamefont{and} \bibinfo{author}{\bibfnamefont{C.}~\bibnamefont{Hess}},
  \bibinfo{journal}{Review of Scientific Instruments}
  \textbf{\bibinfo{volume}{85}}, \bibinfo{eid}{013706} (\bibinfo{year}{2014}),
  \urlprefix\url{http://scitation.aip.org/content/aip/journal/rsi/85/1/10.1063/1.4862817}.

\bibitem[{\citenamefont{Salazar et~al.}(2018)\citenamefont{Salazar, Baumann,
  Hänke, Scheffler, Kühne, Kaiser, Voigtländer, Lindackers, Büchner, and
  Hess}}]{Salazar2018}
\bibinfo{author}{\bibfnamefont{C.}~\bibnamefont{Salazar}},
  \bibinfo{author}{\bibfnamefont{D.}~\bibnamefont{Baumann}},
  \bibinfo{author}{\bibfnamefont{T.}~\bibnamefont{Hänke}},
  \bibinfo{author}{\bibfnamefont{M.}~\bibnamefont{Scheffler}},
  \bibinfo{author}{\bibfnamefont{T.}~\bibnamefont{Kühne}},
  \bibinfo{author}{\bibfnamefont{M.}~\bibnamefont{Kaiser}},
  \bibinfo{author}{\bibfnamefont{R.}~\bibnamefont{Voigtländer}},
  \bibinfo{author}{\bibfnamefont{D.}~\bibnamefont{Lindackers}},
  \bibinfo{author}{\bibfnamefont{B.}~\bibnamefont{Büchner}}, \bibnamefont{and}
  \bibinfo{author}{\bibfnamefont{C.}~\bibnamefont{Hess}},
  \bibinfo{journal}{Review of Scientific Instruments}
  \textbf{\bibinfo{volume}{89}}, \bibinfo{pages}{065104}
  (\bibinfo{year}{2018}).

\end{thebibliography}

% 
% \newpage
% \part*{Supplemental Material}
% 
% \renewcommand{\thefigure}{S\arabic{figure}}
% \setcounter{figure}{0}
% \setcounter{section}{0}
% 
% % \baselineskip12pt
% 
% 
% 
% 
% 
% %\newpage
% \section{Caption for Movie S1}
% \textbf{Differential conductance data.}
% The movie contains two panels, showing the spatial dependence of $\mathrm{d}I/\mathrm{d}U(U_\mathrm{bias})$ in the field of view of the topographic data in Fig.~2b (left) and the Fourier transformation of it (right) for each measured energy $eU_\mathrm{bias}$  at 6.7~K.

\end{document}